\documentclass[12pt]{article}
\usepackage{amsmath}
\usepackage{bm}
\usepackage{amssymb}
\usepackage{natbib}
\usepackage{graphicx}
\usepackage{setspace}
\usepackage{cite}
\usepackage{color}
\usepackage{setspace} 
\usepackage[labelfont=bf,labelsep=period,justification=raggedright]{caption}

\makeatletter
\renewcommand{\@biblabel}[1]{\quad#1.}
\makeatother

\date{}
\defcitealias{gould1977ontogeny}{Gould, '77}
\defcitealias{Hall1999evolutionary}{Hall, '99}
\defcitealias{richardson2002haeckel}{Richardson and Keuck, 2002}
\defcitealias{eldredge1972punctuated}{Eldredge and Gould, '72}
\defcitealias{am1952chemical}{Turing, '52}
\defcitealias{waddington1957strategy}{Waddington, '72}
\defcitealias{glass1973logical}{Glass and Kauffman, '73}
\defcitealias{mjolsness1991connectionist}{Mjolesness et al., '91}
\defcitealias{Salazar-Ciudad:2001lq}{Salazar-ciudad et al., 2001}
\defcitealias{Salazar-Ciudad:2001db}{Salazar-Ciudadet al., 2001}
\defcitealias{goodwin1963temporal}{Goodwin, '63}
\defcitealias{chance1967waveform}{Chance et al., '67}
\defcitealias{cooke1976clock}{Cooke and Zeeman, '76}
\defcitealias{horikawa2006noise}{Horikawa et al., 2006}
\defcitealias{masamizu2006real}{Masamizu et al., 2006}

\begin{document}
\begin{flushleft}
{\Large
\textbf{Evolution-Development Congruence in Pattern Formation Dynamics: Bifurcations in Gene Expression and Regulation of Networks Structures}
}
\\
Takahiro Kohsokabe$^{1}$,
Kunihiko Kaneko$^{2,\ast}$
\\
1 Department of Basic Science, Graduate School of Arts and Sciences, The University of Tokyo, Tokyo, Japan, 153-8902
\\
2 Research Center for Complex Systems Biology, Graduate School of Arts and Sciences, The University of Tokyo, Tokyo, Japan, 153-8902
\\
$\ast$Correspondence to: Kunihiko Kaneko, Research Center for Complex Systems Biology, Graduate School of Arts and Sciences, The University of Tokyo, Tokyo, Japan. E-mail: kaneko@complex.c.u-tokyo.ac.jp

This work was partially
supported by a Grant-in-Aid for Scientific Research (No. 21120004)
on Innovative Areas "Neural creativity for communication" (No.
4103) and the Platform for Dynamic Approaches to Living Systems from
MEXT, Japan.
\end{flushleft}
\section*{Abstract}
Search for possible relationships between phylogeny and ontogeny is one of the most important issues in the field of evolutionary developmental biology.
By representing developmental dynamics of spatially located cells with gene expression dynamics with cell-to-cell interaction under external morphogen gradient, evolved are gene regulation networks under mutation and selection with the fitness to approach a prescribed spatial pattern of expressed genes. For most of thousands of numerical evolution experiments, evolution of pattern over generations and development of pattern by an evolved network exhibit remarkable congruence. 
Here, both the pattern dynamics consist of several epochs to form successive stripe formations between quasi-stationary regimes. In evolution, the regimes are generations needed to hit relevant mutations, while in development, they are due to the emergence of slowly varying expression that controls the pattern change. Successive pattern changes are thus generated, which are regulated by successive combinations of feedback or feedforward regulations under the upstream feedforward network that reads the morphogen gradient. By using a pattern generated by the upstream feedforward network as a boundary condition, downstream networks form later stripe patterns. These epochal changes in development and evolution are represented as same bifurcations in dynamical-systems theory, and this agreement of bifurcations lead to the evolution-development congruences.
Violation of the evolution-development congruence, observed exceptionally, is shown to be originated in alteration of the boundary due to mutation at the upstream feedforward network.
Our results provide a new look on developmental stages, punctuated equilibrium, developmental bottlenecks, and evolutionary acquisition of novelty in morphogenesis.

\section*{Introduction}
The possible relationships between the development of multicellular organisms and their evolution have been the subject of intense research over a century. About 200 years ago, von Baer proposed laws of development, based on observations of development across species, which mainly claimed that the early embryo is mostly conserved across species, while embryonic changes through ontogeny move from a general form common to many species, to species-specific forms \citep{von1828ueber}. Charles Darwin and other biologists of his time interpreted these laws as proof of evolution from a common ancestor \citep{darwin1859origin,muller1869facts,hall2000balfour}. Thus, changes in embryos from a common to a specialized form are regarded as a reflection of evolutional history. Development has been studied in an evolutionary context, and so many biologists have searched for possible relationships between evolution and development. While Ernst Haeckel's claim that 'Ontology recapitulates phylogeny' have been proved incorrect, the search for potential relationships between development and evolution has continued to be of interest to many biologists\citepalias{gould1977ontogeny,Hall1999evolutionary,richardson2002haeckel}.

Previously, this type of research was hindered by a lack of quantitative arguments. To transcend the century-long controversies associated with this research, efforts are being made to quantitatively analyze the evolution-development ('evo-devo') relationship by using gene expression and genome sequence data. In particular, an hourglass hypothesis has been proposed related to the existence of a developmental bottleneck, where differences in gene expressions among several species from the same progenitor decreases at the same developmental stage. This hypothesis suggests that there is a species-wide common stage in development where embryos of different species are similar both in morphology and gene expression patterns \citep{Hazkani-Covo:2005pb,Domazet-Loso:2010cs,Kalinka:2010jt,Irie:2011gd,Quint:2012uk,levin2012developmental,Wang:2013fq}. In spite of these advances, however, a general relationship between evolution and development, as well as its origin, if it exists, remains to be unveiled owing to the limitations in available data on developmental processes along the evolutionary course.

 Species-wide comparison is made using phylogenic trees with branchings to different species, as schematically shown in Figure 1. In contrast, one can make a comparison over species along a single phylogenetic chain from ancestor to offspring. Ontogenies are compared across ancestors along a single evolutionary chain, as shown in Figure 1.
This comparison is hardly possible in practice, as fossil data usually do not include information on developmental processes.
 However, such comparison, if available, gives more straightforward information on relationship between development and evolution, and will provide a basis for species-wide comparison. With such single-chain comparisons, one can gain insight of possible mechanisms that may give evo-devo relationship.

 In contrast to experimental difficulty, however, such single-chain comparison is available by taking advantage of \textit{in-silico} evolution. Indeed, several numerical evolution of developmental process has been recently carried out, by using dynamical-systems models.
These models consider the spatial arrangements and behaviors of cells that are subject to morphogenic gradients and cell-to-cell interactions, and protein expression levels within the cells changes over time by intra-cellular gene expression dynamics. The developmental processes of cellular states are represented by these gene expression dynamics with cell-to-cell interactions to form a spatial pattern of expressions, while the gene regulation networks associated with these dynamics evolve through modifications by genome changes. By establishing a fitness condition to select a specific pattern, the evolution of the network to generate such a pattern can be studied quantitatively. Indeed, with this setup, recent studies of \textit{in silico} evolution have suggested basic mechanisms for stripe formation through development. By establishing conditions to form some number of stripes, two basic modes of gene expression dynamics are revealed, which are generated as a result of feedforward and feedback gene-regulation networks \citep{Salazar-Ciudad:2001lq,Salazar-Ciudad:2001db,Francois:2007rm,Fujimoto:2008jk,Ten-Tusscher:2011vn}. Indeed, these unveiled mechanisms correspond to the two developmental modes in arthropods, i.e., long-term and short-term development, while the detected basic structures in gene regulation networks show some correspondence with those observed in several organisms. So far, however, the relationship between developmental processes and an evolutionary changes has not been explored.

Following former theoretical studies for stripe formation, we focus here on uncovering evo-devo relationship by introducing a fitness condition so that the gene expression of a given output gene in space approaches a prescribed spatial pattern (which is not necessarily periodic in space). Comparing the developmental processes to shape a given gene expression pattern through an evolutionary course under mutation, we found parallelism between evolution and development along the single-chain phylogeny. We name this parallelism as evo-devo congruence, which is observed for the majority of simulation instances. This congruence is based on the correspondence of epochs. In fact, both development and evolution consist of a few epochs that rapidly change to form new stripes, and the slow quasi-stationary regimes between two epochs. Here, drastic change in patterns at each epoch is understood in terms of bifurcation in dynamical systems theory. Both development and evolution adopt the same type of bifurcation to generate epochs that are parallel between evolution and development. Gene regulation networks used to achieve such developmental dynamics are found to consist of a combination of an upstream feedforward network with downstream feedforward or feedback networks. The upstream feedforward network can provide the boundary conditions of same expression levels, through which the temporal oscillation by the downstream feedback network is embedded into a spatial stripe pattern. In rare examples, however, evo-devo correspondence is found to be violated, where change in the upstream feedforward network alters the downstream stripe formation. After first examining extensive numerical results to support the above conclusion, its relevance to biological development and evolution will be discussed.

\section*{Results}

We numerically evolved gene regulatory networks governing development in order to study evo-devo relationship. Here each organism consisted of $M(=96)$ cells aligned in one-dimensional space, where maternal factors were supplied from each end of the space. Each cell had $N(=16)$ genes (proteins) whose expression dynamics were governed by expression levels of other genes through a given gene regulatory network (GRN), while interaction between neighboring cells was mediated via diffusion of expressed proteins. These conditions defined the developmental dynamics of the study. We prepared 100 individuals with slightly modified GRNs. After each gene expression level reached a stationary value through development, we computed fitness from the expression of a prescribed output gene.

 Fitness was defined as the difference between this output expression pattern in space and a prescribed target pattern, with the highest fitness values defined by the best match. We used a genetic algorithm to select the individuals with higher fitness by introducing mutations in the GRN (See Figure 2 for schematic representation and Methods for details).

 Most evolved networks, after few thousand generations, were capable of generating predefined target patterns. An example of the developmental time course to shape such a pattern is given in Figure 3A, where the spacetime diagram of the expression level of the output gene is displayed with the horizontal axis as the developmental time and the vertical axis as the cellular index (i.e., spatial axis). As shown, the target pattern (Figure 3C) is shaped after several developmental stages for stripe formations.(Unless otherwise mentioned, development \textit{after} evolution is plotted for the fittest individual at 2000th generation.)
 Next, we examined how the output gene pattern had evolved, by tracing the final output pattern of the ancestors successively. The output pattern after development of the ancestor at each generation is plotted in Figure 3B, where the color code and spatial axis are identical to those in Figure 3A, while the horizontal axis represents the generation (evolution time) in Figure 3B. The similarity between the developmental (Figure 3A) and evolutionary (Figure 3B) spacetime diagrams are clearly discernible in the Figure 1.

For reference, we have also plotted the developmental course at intermediate (1,300,750,2000) generations in Figure 3D. With successive generations, novel stripes are acquired, moving the system towards the target pattern.

\subsection*{Development with epochs that correspond to those derived through evolution}

Correspondence between developmental and evolutionary spacetime diagrams was commonly observed in our simulations (Figure 3). Additional examples are provided in Figure 4 and in Supplemental Figure S1.

It is remarkable that the pattern formation progressed in a stepwise manner, with respect to both evolution and development. Each stripe emerges not gradually, but discretely at some step in development and in evolution. More interestingly, the correspondence between evolutionary and developmental diagrams is supported by the correspondence of epochs in the two diagrams. This correspondence is valid for a large portion of our simulations. Furthermore, we generally observed good agreement between development and evolution modes, based on the topology of stripe formation: i.e. how later stripes branched from earlier stripes (see Figs.3, 4 and Supplemental Figure S1).

 To quantitatively evaluate the correspondence between evolutionary and developmental spacetime diagrams, we measured the overlap between the diagrams of the output expression levels. The procedure to compute the overlap is shown in Figure 5A. For both diagrams, we took only the temporal regime in which the pattern formation progressed, i.e., we discarded both the early stages where the output gene was not expressed in all cells (i.e., $x_i(t,l)\sim 0$ for all $i$), and the final stage after emergence of the stationary pattern, when no additional changes were observed. The distances between the output expression levels for both the diagrams were then averaged over all spacetime pixels, $\Delta$, thus allowing us to compute the differences between the two diagrams. The distribution of $\Delta$ from approximately 500 evolution trials for different target patterns is shown in Figure 5B, with a peak distribution located at approximately 8\%.

 Note that if the difference between the two diagrams is one stripe over all of the spacetime pixels, $\Delta$ here is evaluated to be 8.2\% (Figure 5B). Hence, the peak value in the distribution is mostly just one stripe difference over all spacetime. Thus for most examples, the spacetime diagrams between developmental and evolutionary processes show remarkable similarity. These results suggest that the correspondence between evolution and development is not an accident, but is a general outcome for most evolution samples. We therefore began an investigation to determine why this evo-devo congruence holds so frequently.

\subsection*{Mechanisms of evo-devo congruence}

\subsection*{i) emergence of slowly changing gene expressions}
 During each epoch, patterns change within a short span for both evolu- tion and development, whereas the pattern remains quasi-stationary between epochs(Figs.3 and 4). Evo-devo congruence is a direct consequence of the congruence of these epochs between evolution and development. Thus, in order to clarify the mechanisms of evo-devo congruence, firstly we must study how such epochal pattern formation is possible for both evolution and development.

 The epoch formation in evolutionary courses is rather trivial. In evolution, the developed pattern does not change until relevant mutations that increase the fitness appear. Until this relevant mutations occur, only neutral mutations are accepted, wherein the patterns are not altered. Hence, the evolutionary course of the developed pattern consists of a quasi-stationary regime and requires several epochs to change the stripe pattern. This epochal pattern change in evolution has been discussed previously using the term ”punctuated equilibrium”\citep{eldredge1972punctuated}.

Correspondingly, we have observed a long quasi-stationary period in development between epochs of rapidly changing stripe patterns; however, this is not self-evident. In this section, we focus on how epochal pattern formation is possible in development. Indeed, we will show that slow gene expression of certain genes is essential to generate a developmental epoch and unveil the origin and function of such slow changes in gene expression.

In Figure  6, we show the change in gene expression at the site marked in Figure  6A, which presents the space-time diagram of the output gene expression. Figure  6B shows the expression dynamics in the marked cells in Figure  6A. Each colored line indicates the expression dynamics of a given gene. As shown, the expression of the output gene (red line) switched between on and off at the time denoted by the crossed red lines in Figure 6A. This switch corresponds to an epoch. As seen in Figure  6B, the gene subset expression changes over time, before and after an epoch. Even though the expression change in most genes, including the output gene, was rather fast to support the epoch, there exists a subset of genes whose expression level changes slowly over time (blue line).

To understand the role of genes with slowly varying expression, a core part of the GRN, which is responsible for stripe formation at each epoch, was extracted (see methods). The core network at each epoch is termed the "working network", as shown in Figure  6C.

 From the working network, it is evident that slowly changing gene expression serves as a control variable for the switch in output gene expression dynamics. The input to the output gene was beyond or below its threshold level at the times marked by crossed red lines in Figure  6A. This change was driven by the on/off switch of gene A, seen in Figure  6C, while the switching in gene A was mainly driven by the slow change in input, denoted by gene S in Figure  6C. The slow change in the expression of gene S served as a parameter cue, providing the timing of the epoch. The morphogen concentration acting on cells was fixed, as indicated by the black horizontal line in Figure  6B. Morphogens initially activated the output gene that provides the first epoch. However, they simultaneously activated gene S, whose expression level increased at a slower rate than that of the output gene. When it exceeded the threshold for the expression level of gene A, the suppression of the output gene was dominant, leading to decreased output gene expression and thereby generating the second epoch. Indeed, the third and fourth epochs were controlled by a slowly varying input in the same manner (data not shown).

We examined several other examples, and found that the working network after evolution always includes a gene with slowly varying expression at the corresponding epoch. The slow gene expression did not give a direct input for the output gene, but gave an input to a gene that gives an input to the target (i.e. gene A and gene B, Figure  6). We then carried out statistical analyses to confirm that control via slow expression changes is a general outcome of evolution.

 In our model, the input term for each expression has a dynamic range given by the threshold ($\theta_i$)and the slope ($\beta$) of the input-output relationship(see method). If the input for a gene is out of the dynamic range (i.e., $-2/\beta<Input-\theta<2/\beta$), the expression of the gene is either 0 or 1. Thus, the time span required for the input to pass through the dynamic range provides an indicator of the time-scale for the control of the input. We computed the timescale of the input to both the output gene and the other genes. In Figure  7, the evolutional change of these two time scales are plotted (the latter time scale is the slowest change among all genes that have a path to the output gene). As shown, the time scales are nearly equal at the first generation, indicating the absence of slow expression control. Throughout evolution, the timescale of the output gene was not altered, which also supported the notion of epochs with a short time span of change. In contrast, the time scales of other genes slightly increased, such that the ratio of the target timescale to that of others decreased, reaching 1/5 of the average. Thus the relative timescale of input gene expression to the change in target expression was slower during evolution. Hence the results of Figure  7 support the emergence of epochs by slow genetic control. 

\subsubsection*{The origin of slowness in expression}
Questions remained regarding the origins of such slow expression dynamics. Following analysis of all examples, we concluded that the origin of slowness could be attributed to the following two mechanisms:\newline
\newline
(i) The existence of genes with small rate constants $\gamma_i$ associated with expression change. The expression dynamics in our model includes a parameter $1/\gamma_i$, representing the time constant for change. Hence if some gene $i$ has a small $\gamma_i$ value, expression changes slowly. While this itself may appear rather trivial, it should be noted that the rate parameters $\gamma_i$'s after evolution are distributed by gene $i$, and some genes have smaller $\gamma_i$ values. Therefore, the expression levels of genes with small $\gamma_i$ values function as a slow variable. Indeed, in the example presented in Figure 6, $\gamma_i$ for the gene $S$ is 0.063; a full order of magnitude smaller than the others. Through evolution, genes with distinctively small $\gamma_i$ values appear, even though we initially established nearly uniform $\gamma_i$ values for all genes.\newline
\newline
(ii) Expression levels near the threshold. The expression dynamics here have a threshold $\theta_i$. If the input to the gene is larger (or smaller) than $\theta_i$, it is expressed (or suppressed), respectively. When the input term from other genes to the gene $i$ is close to $\theta_i$, then, the expression level can be balanced at an intermediate value between 0 and 1. Indeed, if the deviation of input from $\theta_i$ is smaller than $1/\beta$, the inverse of sensitivity, then the expression level of $x_i(l,t)$ is no longer attracted to $0$ or $1$. In this case, this stationary state is less stable than those closer to 0 or 1 (see Supporting text S1 for the mathematical explanation using the Jacobian matrix). Hence, the time-scale around this fixed point is longer in duration.\newline

This slow relaxation to the stationary state as a single-cell dynamics is extended through the entire space, mediated by the diffusion interactions with other cells. With diffusion, the slow expression change of a certain cell can propagate spatially to other sites, to change their expression levels slowly.

\subsection*{ii) mechanisms for pattern formation and their dependency}
Now, we show how stripes(valleys) are formed in developmental process here, based on gene regulation dynamics, cell-to-cell diffusion, and morphogen gradient.
Through extensive analysis of 500 samples of the evolved pattern-formation, we confirmed that the stripe formation process is reduced to only two basic mechanisms in gene expression dynamics with corresponding GRN structures. In fact, these two mechanisms have previously been identified and studied extensively, which are known as feedforward and feedback regulations\citep{Salazar-Ciudad:2001lq,Salazar-Ciudad:2001db,alon2006introduction,Francois:2007rm,Fujimoto:2008jk,cotterell2010atlas,Ten-Tusscher:2011vn}.

\subsubsection*{Feedforward}
The classic mechanism for stripe formation, which was analyzed in the segmentation process in Drosophila, is feedforward regulation. This mechanism has been analyzed both theoretically and experimentally\citep{von2000segment, jaeger2004dynamic, ishihara2005cross}. In this case, a gene 'reads' the morphogen gradient for spatial information, to establish an 'on/off' response under a given threshold level, so that the gene is expressed on the one side of space, and non-expressed on the other side. Another 'downstream' gene receives positive (or negative) input from this gene, and negative (or positive) input from the morphogen, then responds to create another segmentation in space, if the threshold parameters satisfy a suitable condition. By combining this feedforward regulation, more stripes are formed for the downstream gene. The corresponding GRN does not require feedback regulation, or cell-to-cell interaction by diffusion; only unidirectional, feedforward regulation from morphogen input to downstream genes is required. This feedforward regulation frequently exists in our evolved GRN, and is used to generate at least some stripes.
\subsubsection*{Feedback oscillation within a boundary}
The other mechanism for stripe formation, commonly observed in the present simulations, takes advantage of feedback regulation to produce a temporal oscillation in the expression level. This temporal oscillation at a single-cell level is then fixed into a spatial periodic pattern by the diffusion among cells. A typical core network structure and expression dynamics are shown in Figure 8A. Here, gene A activates the expression of both itself and gene B, while gene B suppresses the expression of gene A. Since this network is just a typical negative feedback loop, it produces a temporally oscillating expression when the parameter values for the expression dynamics are appropriate. Now, with the diffusion of B under an appropriate boundary condition, this temporal oscillation is fixed into a spatially periodic pattern (see Figure 8B).

 Consider a case where the input from the morphogen M suppresses the activation of A at the boundary. Again, without input from M at the boundary, there appears temporal oscillation in the expression of A and B. At this boundary, the expression of gene B is also suppressed. Then the protein B at the adjacent upper site diffuses to this boundary. Subsequently, the expression of B at the site is decreased, so that the suppression of A is relaxed(Figure 8B, bottom). Then the protein of gene A is fixed to a higher level, instead of oscillating. This also leads to an increase in the expression of B. At the next upper sites, oscillation still remains. When the expression of A is low, the diffusion of protein B from the lower site suppresses the activation of A, so that the increase of the expression of A no longer occurs. Thus, the expression level of A is fixed at a lower level. With this process, temporal oscillation of one period is mapped into one spatial stripe. The same fixation process is repeated with the subsequent oscillation at further upper sites, since at the nearest lower site, the expression of gene A is fixed to a lower level. Thus, the temporal oscillation is recursively fixed to a spatially periodic pattern. With this mechanism, the stripe pattern in space is formed and fixed (For detailed theoretical analysis, see Supporting Figures S2, S3).

This mechanism is analogous to the classic Turing pattern in which case the suppression of B to itself is necessary to exclude a spatially homogeneous, temporally oscillating state. Here, the diffusion of the inhibitor gene B works in the same way as the Turing pattern\citepalias{am1952chemical}, but the mechanism here adopts temporally oscillatory dynamics, and is understood as Turing-Hopf bifurcation\citep{PhysRevE.54.261}. In an interacting 2-cell system, this differentiation from the oscillatory state is understood as, a saddle node bifurcation on invariant cycle (SNIC). SNIC in a globally coupled dynamical system has been studied as a mechanism for cell differentiation from a stem cell\citep{PhysRevE.88.032718}.

We note two points. The mechanism here uses the suppression at a boundary to fix a pattern, instead of the inhibition of B to itself, and thus the boundary condition is important. Next, the mechanism also resembles a classic wavefront mechanism, but is different. In the wavefront mechanism, the temporal oscillation is fixed into a spatial periodic pattern through input from the morphogen gradient and the growth of the system size. In the mechanism, however, diffusion (or cell-to-cell interaction) is not essential, and external manipulation by the morphogen gradient for all cells leads to fixation forming a stripe from the oscillation. In our case, such external manipulation exists only at the boundary, and further stripe formation progresses spontaneously by the diffusive cell-to-cell interaction.

\subsubsection*{Stripe formation order by the combination of the two mechanisms}

All of the potential evolved stripe formation processes in our model could be generated by a combination out of four possible ways of combining these two developmental mechanisms, sequentially. However, for the feedback mechanism to work, the boundary depending on the morphogen has to be established in advance, to fix the temporal oscillations to spatial stripes. Thus, the feedforward mechanism to read the external morphogen is needed to produce the boundary. Otherwise, no stripe will be formed, so that such networks will not remain in the evolutionary simulation. Hence, we consider only two combinations: feedforward-feedforward and feedforward-feedback. Indeed, these two cases are the bases for all of the possible developmental processes evolved in our simulation.

\subsubsection*{Sequential Feedforward mechanisms}
Stripe formations involving the combination of feedforward mechanisms have been extensively studied. In some examples, the developmental processes evolved here are achieved by sequentially combining feedforward processes, where cell-to cell interaction is not needed. Consider a new feedforward mechanism, added at some point downstream from an upstream feedforward circuit. As long as the upstream mechanism is not affected by the downstream mechanism (which is true if there is no feedback from the latter to the former), the stripe formation progresses first by the upstream mechanism, and then, at a later epoch, the stripe is generated by the downstream mechanism. This ordering in the developmental process agrees with the order of evolution, since the downstream mechanism is acquired later in the evolutionary course. Hence, in this simple sequential feedforward mechanism, the evo-devo correspondence is a natural outcome. Since the evolved GRN typically has slower gene expression dynamics that control the downstream expression as already shown, the stripe formation will occur sequentially in developmental time, with some delay, in agreement with evolutionary time course. 

The evolved network illustrated in Figure 9A consists of a combination of sequential feedforward networks and a downstream feedback network. Through evolution, first the feedforward network via gene 10 and gene 3 (see Figure  9D) is acquired at around the 10th generation. Then, a domain in the middle space is shaped in development as shown in Figs. 9B and 9E. Later, at the 88th generation, another feedforward network via gene 12 is attached downstream through evolution (Figure 9F). With this attached network, a domain is shaped in the interior of the earlier domain as seen in Figs.9F and 9G, right after the earlier domain formation is shaped. The shaping of domains is successfully completed at an early stage of development. This leads to the evo-devo congruence. Later, this modified domain in the middle works as a boundary condition for the subsequent feedback network to be discussed.

\subsubsection*{Feedback-oscillation mechanism attached downstream of the feedforward network}

The Upstream feedforward network  is indeed necessary for the feedback mechanism to work as already explained. In development, the stripe formation by the feedback mechanism cannot work without a boundary, and only after the appropriate boundary condition is generated by the feedforward mechanism. On the other hand, the feedforward circuit is first acquired in the earlier stage of evolution to increase fitness, and later the feedback-oscillation is obtained to create further stripes using the former feedforward stripe as a boundary. Hence, we again observed good agreement in the time courses of stripe formation development and evolution, as long as the latter feedback mechanism does not influence the former feedforward mechanism. Slower expression change of controlling genes as already discussed works for the separation of two epochs.

An example of evo-devo congruence caused by feedback-oscillation downstream of the feedforward mechanism is displayed in Figure 9A. Corresponding space-time diagrams of evolution and development are presented in Figs.9B and 9C. Evo-devo congruence is detected, in particular between the third and fourth upper stripes. These two stripes are generated by the oscillation-fixation mechanism generated by the feedback loop (Figure 9A), attached downstream of gene 3, which is a component of the feedforward network from a maternal morphogen. This feedback module is inhibited by two morphogens and gene 5, so that this oscillation does not start without an input for activation. The only activation input for this feedback module is gene 3, which is expressed only in a domain restricted by the upstream feedforward network. Thus, the oscillation starts after the expression level of the gene 3 is sufficiently high (Figure 9I), and thus is bounded within the domain, maintaining the expression of gene 3 (Figure 9H). Following the mechanism discussed in the next section, a stripe is generated in this domain. This feedback oscillation is regulated by the upstream feedforward network but does not disturb upstream feedforward expression.

\subsection*{iii) Parallelism between the working GRNs of evolution and development}
The results in the last section suggest that the ordering of working networks over epochs is in agreement with development and evolution, and that both progress from feedforward-based networks to networks including feedback loops in addition to feedforward networks. We examined the validity of this ordering statistically by using all the data in our simulated evolutions.

We first examined whether a working network includes a feedback loop, and computed the fraction of purely feedforward networks that do not include a feedback loop at each epoch. 
As shown in Figure  10A, this feedforward ratio is close to 1 at the first epoch ($\sim 0.85$), and it decreases in later evolutionary epochs.
Nevertheless, as the working network size for the $i$-th epoch $k_i$ increases, the probability of a network without feedback loops is expected to decrease. This probabilistic decrease is estimated as $f^{k_i/k_1}$, which is also shown in Figure 10A. The observed rate decrease without feedback loops is much higher than this estimate. Hence, the fraction of feedforward networks is significantly higher during the first epoch, while the fraction with feedback loops increases faster than was estimated by using the increase in network size. In comparison, we also plotted the fraction of pure feed-forward networks from random networks of corresponding size (Network size were computed after randomly generated networks were processed such that genes that was not included in a path from the morphogen to the target gene were removed). The fraction in the evolved network was much larger. Thus, the feed-forward network was preferentially selected.

We also checked whether the ancestral network is conserved by computing the fraction of networks preserved in later epochs (see methods). 
This overlap ratio is shown in Figure 11A as a function of evolutionary epochs. Values were higher than 0.75 over the epochs, such that over 75\% of the ancestral working networks were preserved during evolution. In summary, the results shown in Figure 10A and 11A indicate that working networks mostly consisted of feedforward networks, and were well conserved throughout evolution. Additionally, later in evolution, feedback loops were added.

The fraction of pure feed-forward networks in the working network is plotted against the {\em developmental} epochs in Figure  10B. The network size was large at the first epoch, such that the ratio at the first epoch was small. Still, the decay of the fraction is much larger than expected by the probability calculations due to the increase in network size.

Finally, we checked the overlap ratio of the working networks between evolution and development. As shown in Figure  11B, the overlap remains high throughout the epochs, indicating that the working networks in evolution and development correspond with each other. Thus, the same pattern formation dynamics are adopted in the same order between evolution and development.

\subsection*{iv) A slowly varying expression level works as a bifurcation parameter to produce a developmental epoch
}
So far we have uncovered the existence of slow expression change working as a control for the output gene expression and the combination of feedforward and feedback networks. These are important for evo-devo correspondence in pattern formation and gene-expression ordering, but we need to understand how these two features lead to fast switch-like change in target patterns at epochs, and how these are correlated in development and evolution. Here we describe this congruence of fast switching dynamics in evolution and development, in terms of bifurcation in dynamical-systems theory.

Consider the example of the network presented in Figure 6, with a slowly changing expression of Gene S. When the expression level of gene S (slow variable) increases slowly and reaches a certain level, the expression level of gene A increases from $\sim0$ to $\sim1$. Input changes to gene B may then lead to bifurcation. Here the morphogen (gene M) activates gene S and gene B, while gene S activates gene A, and gene A subsequently inhibits gene B. If the expression level of gene S is smaller than the total activation input to gene B, the dynamics of expressions of gene A and B are given by the flow presented in Figure 12 (upper left). The nullcline of gene B forms z-like structure in this phase space, which crosses the perpendicular nullcline of gene A, at coordinates near (0,1). As the expression level of gene S increases, the nullcline for the expression of gene A moves horizontally, so that the fixed point at $(x_A, x_B)\sim(1,0)$ disappears and is replaced by the fixed point at $(x_A, x_B)\sim(0,1)$, as seen in Figure 12 (bottom left). Thus, the bifurcation between fixed point attractors occurs with a change in the expression level of gene S as the bifurcation parameter.

\subsubsection*{Bifurcation behind Evo-Devo correspondence}
During this study, we observed that the developmental process consisted of a quasi-stationary regime and epochs to form new stripes, due to bifurcations resulting from slowly changing expressions as parameters. This is relevant to achieving evo-devo congruence, since the evolutionary process also consists of a quasi-stationary regime prior to the emergence of a relevant mutation capable of increasing fitness within a relatively short time span. Indeed, such mutations change the gene expression dynamics drastically to form a new stripe, which again is regarded as a bifurcation.
 At a certain generation in the evolution, a mutation occurs to add an inhibition path from gene S to gene A (Figure 12). This mutation occurs in a discrete manner: Whether a path exists or not, it is not represented as a continuous change in a parameter value. However, we can introduce a continuous strength parameter that changes from $0$ to $\pm1$, which can be regarded as a bifurcation parameter. Then with this continuous change, an on/off discrete change appears at a certain value of path strength that depends on the threshold of gene A. Dynamics of the expression of gene A and B are represented in the two-dimensional state space in Figure 12(right column). At a lower strength in the path, the nullcline of gene B expression changes so that the former stable fixed$(1,0)$ point exhibits a saddle-node bifurcation, to move to another fixed point (0,1). Hence, the mutational change in the network leads to a bifurcation. As seen in Figure  12, this bifurcation through the evolutionary process agrees with that observed during development.

 After examining hundreds of numerical evolution simulations, the results were summarized as follows:
\textit{Development}: slow change during expression works as a bifurcation parameter, and bifurcation in the expression dynamics generates a novel state, which gives rise to an epoch.
\textit{Evolution}: search for mutation resulting in relevant change to a new state. Epoch in evolution is also generated by the same bifurcation. In this way, evo-devo correspondence is achieved through bifurcation.

\subsection*{Violation of Evo-Devo correspondence}
 Although evo-devo correspondence was frequently observed and was discussed as a natural outcome of the combination of network motifs for development, small, but non-negligible portions of the simulation runs exhibited deviation from this evo-devo correspondence. An example of such an exception is presented in Figure 13A (See also Supporting Figures S4 and S5 for additional examples).
In this example, the developmental and evolutionary diagrams differ distinctly, not only in the timing of the formation of the second and third upper stripes, but also in the topology in their branching.
During the course of evolution, there is a drastic change in the final pattern, at approximately 1272-1273 generations. Here, only a single mutation occurred in a GRN (addition of a single path). In this example, the feedback oscillation of gene 5 was responsible for the output gene expression, in particular for the second and third stripes, while the expression of gene 6, which lies upstream of gene 5, acted as a boundary for the feedback oscillation, which also contributed to the expression of the output gene.
 In Figure 13B, the gene expression dynamics of the selected genes 5 and 6, as well as the output gene, are displayed for generations before and after this mutation, in the left and right rows, respectively. Here, the mutation occurred upstream of gene 6, and reduced the range in which the gene was expressed, accordingly. The expression around sites 60-70 was subsequently suppressed, allowing for the formation of an additional stripe near site 70, while at lower sites (around site 60) the expression level continued to oscillate, forming a stripe much later. Hence the temporal ordering of the formation of the second (near site 80) stripe, and that of the third (near site 70) stripe was reversed by this mutation. Indeed, before the mutation, the third and fourth stripes were generated together (while the second stripe did not exist), and after this mutation, the second and fourth stripes were generated together, and the third stripe was shaped later. Thus, the ordering and topological branchings of stripes were altered by the mutation, which led to a violation of the evo-devo correspondence.

To summarize, the violation of the correspondence was due to an upstream expression change resulting from mutation, which caused a change in the boundary condition for the feedback oscillation of the downstream expression gene. We have studied several other examples that showed a violation of evo-devo correspondence, and confirmed that differences in the topology in stripe branchings between development and evolution is caused by mutation upstream of the feedforward mechanism acting as a boundary of the feedback mechanism (See Supplementary information for additional examples).

\section*{Discussion}
\subsection*{Summary:}

 The potential relationship between phenotypic dynamics to shape phenotypes and evolution in genotypes has been the focus of the evo-devo field, since the time when genetic assimilation was first proposed by Waddington \citepalias{waddington1957strategy}. The relationship has been investigated in RNA evolution \citep{ancel2000plasticity} and gene expression dynamics \citep{ciliberti2007robustness,kaneko2007evolution}, as also summarized in recent reviews \citep{wagner2005robustness,kaneko2006life,soyer2012evolutionary}. Still, studies to establish relationships between multicellular pattern formation dynamics and evolutionary processes that shape the pattern remain premature both in theory and experimentally.

 Here, we carried out extensive simulations to evolve gene regulatory networks subject to fitness requirements, in order to generate a predefined target pattern for the expression of a given output gene. The main results of the present paper are summarized as follows:
\subsubsection*{1: Epochs of development as bifurcation:}
 The developmental course of the expressions of the output gene, after evolution, consisted of a few epochs characterized by rapid temporal change in gene expression and a quasi-stationary regime with slow temporal change between the epochs. The slow quasi-stationary regime is due to expression levels of some genes that vary slowly over time, while the rapid drastic change is due to a bifurcation in the expression dynamics. The slowly varying expressions emerge as a result of evolution, and they work as {\sl bifurcation parameters} to control the fast change in the expression of the output gene.
\subsubsection*{2: Punctuated equilibrium in the evolution of morphology as bifurcation:}
 Likewise, the evolutionary course of expression dynamics consists of a few epochs with a drastic change, and a quasi-stationary regime between the epochs. The drastic change is again represented by a bifurcation, which is caused by mutations in the gene regulation network.
\subsubsection*{3. Evo-devo congruence through common bifurcations:}
 In most cases, we observed good correspondence between development and evolution, with regard to spatiotemporal dynamics from a uniform state to a target pattern. We observed good agreement between development and evolution when evaluating epoch changes from one pattern to another, as well as the ordering of epochs. Indeed, the same bifurcations occurred for both, and thus the evo-devo congruence was due to the common bifurcation at each epoch.
\subsubsection*{4. The combination of feedforward and feedback gene regulation networks to support developmental epochs:}
 The combination of feedforward and feedback modules in gene regulation networks provides successive bifurcations at epochs. The upstream feedforward network converted the external gradient of the maternal factor into an output pattern, while the feedback loop converted the temporal oscillation of gene expression into a spatial stripe, under a given boundary condition provided by the feedforward expression dynamics. The evo-devo correspondence was preserved as long as the upstream feedforward network was maintained.

\subsubsection*{5. Violation of evo-devo correspondence through modification of upstream feedforward regulation under downstream feedback mechanism:}

 In rare examples of our simulated runs, we observed violations of evo-devo congruence. These violations were always associated with a structure of the upstream feedforward network and a downstream feedback loop, in which modification of the upstream feedforward network changed the boundary condition of the downstream feedback.

 This then raised questions as to why the sequential feedforward network was excluded therein, and whether the violation always involved the feedforward-feedback combination.
The feedforward mechanism reads the morphogen gradient of a maternal factor, so that the feedforward-feedforward process transfers spatial information of the maternal gradient sequentially, from upstream to downstream. The flow of spatial information was unidirectional, so that the downstream genes could not generate new stripes on their own. Since each stripe location was defined by the expression of the upstream genes, the downstream genes could not translate their stripe location in parallel.
For the violation of evo-devo congruence to occur without the loss of fitness, two mutations, one to delete a stripe and one to add a stripe, had to occur at the same time, otherwise, downstream stripe formation would be damaged, and fitness would decrease. As two such simultaneous mutations are less probable, the violation of evo-devo congruence under feedforward-feedforward network rarely occurred. Conversely, in the case of the feedforward-feedback network, the downstream feedback loop maintained the stripe formation mechanism, and the upstream changes affected only the boundary condition. Hence, as a result of a single mutation, the stripe position could be shifted without destroying it. In this instance, only a single mutation was needed, which is why the violation of evo-devo congruence we observed was always in association with the feedforward-feedback network rather than through a sequential feedforward network.

\subsubsection*{Relevance of our results to developmental and evolutionary biology}
 Now we discuss the relevance of our results to evolution and development of biological patterns, corresponding to the points noted above.\newline
\newline
(\textbf{1}) Note that the developmental process evolved in our simulation involved slow change in concentrations of some chemical controlling the dynamics. Slow gradual changes in the concentrations of several chemicals are known to play an important role in the developmental process, which may involve some signal molecules, hormones, and morphogens \citep{carroll2009dna}.
The developmental process is generally believed to consist of successive stages, each of which involves time spans with slow gradual change, and epochs involving drastic change leading to the next stage. Novel dynamical processes are necessary for such epochal changes\citep{carroll2009dna}. This empirical facts in development are consistent with our observations, while our bifurcation scheme provided an interpretation for commonly observed developmental stages.
Because processes that generate such drastic changes are not fully understood so far in developmental biology, it will be relevant to analyze such changes in terms of dynamical systems, in particular, by bifurcation against slow change in some concentration of chemicals.\newline
\newline
(\textbf{2}) The existence of a quasi-stationary regime and rapid change are often discussed in evolution in terms of punctuated equilibrium \citepalias{eldredge1972punctuated}. Indeed morphological changes observed through fossil data have suggested these temporal modes throughout the course of evolution. Our results suggest that such temporal modes can be explained as bifurcation. Indeed, research has suggested that novel developmental events are acquired in evolution as a result of bifurcations (i.e., \textit{evolution as bifurcation})\citep{Francois:2012yg, jaeger2012inheritance}.

From fossil data it is difficult to confirm our theoretical consequence that the acquisition of morphological novelty in evolution is achieved by bifurcation in developmental dynamics. Alternatively, we expect that  a novel morphological pattern may be achieved by imposing suitable changes in gene expression dynamics that might correspond to evolutionary change. For example, by introducing a hormone and over-expressing a single gene, Freitas et al succeeded in inducing fin distal expansion and fin fold reduction in zebrafish, which conceivably represented a prototype of vertebrate appendages\citep{freitas2012hoxd13}. The induced epigenetic change leads to a novel gene expression pattern, thereby generating a stripe pattern. This may correspond to bifurcation of a spatial pattern due to genetic change of expression dynamics in our study.\newline
\newline
(\textbf{3}) In our study, correspondence between evolution and development is achieved through common bifurcations. It is difficult to check this correspondence directly from experimental data since the morphology is not easily traced through an evolutionary course, while the comparison of phylogeny and ontogeny usually involves examination of the morphology only of present organisms that have diverged from common ancestral species(See Figure  1). Hence, it is not possible to directly confirm our evo-devo correspondence. However, if the morphological novelty is a result of bifurcation, different novel morphologies are expected to be diverged from a common ancestral pattern, through different bifurcations. This viewpoint is consistent with von Baer's third law of embryology, which claims that a common basic morphological feature of the group emerges in advance of special features for each species. If we assume that ancestral features are basic for the group, our result suggests that von Baer's third law is due to morphological constraint and bifurcation of developmental processes induced by genetic change.

Currently, a popular topic in the evo-devo field is to examine the existence of phylotypic stages and the validity of the developmental hourglass \citep{Hazkani-Covo:2005pb}. Several recent studies have investigated these topics for different species including Drosophila \citep{Kalinka:2010jt}, vertebrates \citep{Irie:2011gd}, plants\citep{Quint:2012uk}, Caenorhabditis \citep{Quint:2012uk} and the soft-shell turtles \citep{Wang:2013fq}.
As previously mentioned, our study was unable to provide direct evidence to support the developmental hourglass, since it was not intended for species-wide comparison but for a single chain in the phylogeny to unveil the basic mechanism for the evo-devo congruence. Also, all individuals in our model were subject to the same initial conditions, with suppressed expression of genes under the same external morphogen gradient, which could not adequately mimic the conditions that would be observed for real multicellular organisms \citep{sander2004evo}. Despite these deficiencies, the results of this study may have relevance to species-wide comparison also. Our results suggest that evolutionary branching from common ancestral pattern to generate diverse morphological patterns occurs through bifurcation in dynamical systems from common ancestral pattern. Diversification from the bottleneck in evolution and development can be understood accordingly, which may give a basis for the hourglass model.\newline
\newline
(\textbf{4}) The importance of feedforward and feedback regulations in development has now gained more extensive recognition. The relevance of the successive combination of feedforward networks has been recognized in long-germ segmentation processes in Drosophila, together with theoretical analyses  \citep{von2000segment, jaeger2004dynamic, ishihara2005cross}.
On the other hand, the relevance of a feedback loop to form temporal oscillations has been revealed for several decades \citepalias{goodwin1963temporal,chance1967waveform,cooke1976clock,horikawa2006noise,masamizu2006real}. In vertebrates, Pourqui\'e discovered that somite genesis is achieved by mapping this temporal oscillation to a spatial stripe formation, where a wavefront model is applicable \citep{pourquie2003segmentation}. Our mechanism to fix the temporal oscillation to the spatial pattern is similar to the wavefront model, but has some differences. In the wavefront model, a combination of the morphogen gradient, size growth, and oscillatory dynamics forms the stripe pattern, while, in our case, a combination of cell-to-cell interaction with diffusion and oscillation leads to stripe formation under the boundary conditions provided by upstream feedforward gene regulation. This distinction will be experimentally verifiable by determining whether the cell-to-cell interaction is essential to stripe formation.

Here we also demonstrated the importance of the feedforward-feedback combination, to read external morphogen information leading to robust stripe formation in a bounded domain, where the boundary condition to determine the domain is supported by the upstream feedforward network. Complex gene regulatory networks in the present organisms often include a combination of feedforward-feedback networks \citep{carroll2009dna}, although their functional roles have not been fully uncovered. It will be important to elucidate the role of the feedforward network as a boundary-maker and the role of the feedback loop in robust patterning, as suggested here.\newline
\newline
(\textbf{5}) Experimental confirmation of the violation of evo-devo congruence through modification of the upstream feedforward network, with a conserved downstream feedback loop, is expected to be difficult, considering limitations in available evolutionary data. Therefore, we propose to examine whether morphological novelty arises as a result of modification of upstream feedforward regulation under feedback regulation. While direct examination of our feedforward-feedback hypothesis from the data is difficult, it is possible to evaluate the hypothesis by externally destroying  the upstream feedforward network while retaining the feedback loop.

\subsection*{Future Issues}
The present study is an initial step towards resolving the larger issue of evo-devo correspondence. Even within the present model, a number of issues remain to be clarified, as follows:\newline
\newline
(i) Even though we have confirmed that our result is independent of the details in the model, such as the cell number, gene number, model parameters, and the form of the external morphogen profile, further study is necessary to confirm the universal applicability of our results.\newline
\newline
(ii) In the gene expression dynamics after the evolution, we found that there always exists a slowly varying gene expression level that works as a control parameter. Thus far, we have not uncovered the conditions responsible for the emergence of this slower mode, which controls other expressions that are relevant to fitness. By modifying coupling with this slower mode, the output behavior that it controls is more readily altered, so that evolution can be facilitated. Therefore, the emergence of slowly varying expression(s) may be evolutionary advantageous. It is important to investigate the generality of the emergence of this slow variable.\newline
\newline
(iii)We have not observed the classic Turing mechanism \citepalias{am1952chemical} in the developmental process by evolved networks. Under the influence of a morphogen gradient, it may be natural to use the maternal information effectively with respect to evolution. It is then an open question whether without the external information (but by imposing the boundary condition instead), the Turing-pattern mechanism can evolve dominantly.\newline
\newline
(iv)How do evolutionary reachability of the target and complexity in the developmental process depend on the predefined target pattern? It may be expected that as the target pattern is more complex, it takes more time to evolve GRN to produce such patterns, and development involves more epochs, but is there a way to quantitatively characterize such complexity?\newline
\newline
(v) We have explained evolution-development congruence in terms of the correspondence of epochs, but it is not clear whether quantitative congruence exists beyond this level. For example, does the time span for the quasi-stationary regime between two epochs correlate between development and evolution? In other words, if the evolutionary search time to generate a relevant mutation for the next epoch is longer, then, is the quasi-stationary regime before the epoch also longer in development? Our preliminary results suggest that this correlation exists for cases where small $\Delta$ values are observed, while further analysis is required to clarify the conditions and mechanisms for such congruence.\newline
\newline
(vi)Extension of our model for higher spatial dimensions, introduction of size (cell number) growth through development, and inclusion of recombination in a genetic algorithm will be important in future studies. Furthermore, it is important to note that real morphogenesis in multi-cellular organisms is far more complex than these models, and is not necessarily governed only by the reaction-diffusion mechanism. Cell rearrangement under mechanical stress could also play an important role, and inclusion of development mechanisms will be important. Still, we also note that previous research has indicated that macroscopically represented, stress-induced pattern formation can also be represented by equations of the reaction-diffusion type \citep{murray2002mathematical}. Hence, the present conclusions on evo-devo congruence through bifurcations may be applicable beyond development based on a reaction-diffusion system.\newline
\newline
(vii)Last but not least, the implications of our single-chain-phylogeny study on species-wide comparison have to be explored. For example, by adding population division and speciation process with imposing different target pattern to the model, species-wide extension will be available. That future study will be important not only for the validation of our results, but also for further understanding of evo-devo relationship in species-wide comparison.

\subsection*{Concluding remark}

In contrast to recent advances in experiments aimed towards analyzing the evo-devo relationship at a quantitative level, theoretical studies based on dynamical systems and statistical physics are still in their infancy. While we acknowledge that our current model may be oversimplified, we hope that the present work can act as a springboard to launch future cooperative efforts in the field of evo-devo between theories and experiments.

\section*{Methods}
\subsubsection*{Gene Regulation Network(GRN) model for pattern formation}

A cell's state is represented by the expression levels of k genes/proteins, $x_i(l,t)$, involving the protein expression levels of the $i$-th gene in the $l$-th cell at time $t$, representing $N$ genes ($i=1,.,N$) and $M$ cells, aligned in a one-dimensional space. A protein expressed from each gene either activates, inhibits, or does not influence, the expression of other genes, in addition to itself. For simplicity, we assumed that the change in the $i$-th protein expression level is given by the equation:
\begin{equation}
\frac{\partial x_i(l,t)}{\partial t}=\gamma_i(F(i,l,t))-x_i(l,t))+D_i\frac{\partial^2x_i(l,t)}{\partial l^2}
\end{equation}

with

\begin{equation}
F(i,l,t)=f(\sum_jJ_{i,j}x_j(l,t)-\theta_i)
\end{equation}

where the term $-x_i(l,t)$ in (1) provides a measure of the degradation of the $i$-th protein with $\gamma_i$ as its rate \citepalias{glass1973logical,mjolsness1991connectionist,Salazar-Ciudad:2001lq,Salazar-Ciudad:2001db}. The expression level is scaled so that the maximum level is unity. The function $f(x)$ is similar to a step function, where the function approaches 1 as $x$ is increased to a positive side, and approaches 0 as $x_i(l,t)$ is decreased to a negative side: In other words, if the term $\sum_jJ_{i,j}x_j(l,t)$ is sufficiently larger than the threshold $\theta_i$, then $F(i,l,t)\sim1$, which indicates that the gene is fully expressed, and if it is smaller than the $\theta_i$, then $F(i,l,t)\sim0$, which indicates that the gene expression is suppressed. Here, we chose, $f(x)=1/(1+e^{-\beta x})$, where $\beta$, which was set to 40, denoting the sensitivity of the expression at the threshold. Roughly speaking, it is proportional to the Hill coefficient.

The gene regulation network was introduced to our model based on work reported in earlier studies. In Figure 2, each node of the network represents a gene, and the edge of the network represents the interaction between genes, given by $N$$\times$$N$ matrix $\bm{J}=\{J_{i,j}\}$: where $J_{i,j}$ is $1$, if gene $j$ activates the expression of the gene $i$, $-1$ if it suppresses the expression, and $0$ if there is no connection. All cells have an identical regulatory network, with the same parameter values, which are determined by genetic sequence in the genome.

Finally, the last term in Eq. (1) shows the diffusion of a protein, between neighboring cells, with $D_i$ as the diffusion constant. For the majority of the simulations described here, we set $M=96$, and $N=16$, while preliminary simulations adopting larger values for these did not alter the conclusion in the present paper.

\subsubsection*{Initial/boundary condition}

As an initial condition, the expression levels of all genes were set to 0. Furthermore, external morphogens, which are denoted as the proteins 0 and 1, are supplied externally. Fixed linear morphogens are induced from both sides for cellular use, so that $x_0(l,t)=x_0(l)=(M-l)/M$ and $x_1(l,t)=x_1(l)=l/M$. We also evaluated a case involving a gradient with an exponential dependence in space, as Cexp($-l/\xi$), but this condition did not alter the conclusions presented in this study. Discrete Neumann boundary conditions were adopted at both ends for this study, i.e., $x(1)=x(2)$ and $x(M)=x(M-1)$.

\subsubsection*{Definition of fitness}

To study the evolution of morphogenesis, we imposed a fitness condition to generate a given specific target pattern, for the expression of a given output gene. By setting a target pattern as $T(l)$, the fitness $f_i$ was defined as the sum of the distance between this target pattern and output gene expression at each cell, as:
\begin{equation}
f_i=\sum_l1-|T(l)-x_{output(l)})|
\end{equation}
where $l$ is a cellular index.
From the equation, the smaller the distance, the higher the fitness. Here the output gene pattern was defined after a given transient time, which was chosen to be large enough to reach a stationary pattern. For each genotype (i.e., GRN and a set of parameter values), the fitness was thus computed, after simulating each set of pattern dynamics.

For this analysis, we chose 100 individuals with different genotypes. Among these individuals, those who had higher fitness values were preferred to be selected for the next generation. Selection of the individual $i$ with a fitness value of $f_i$ was defined by
\begin{equation}
p_i=\frac{e^{f_i}}{\sum_k e^{f_k}}
\end{equation}
The denominator summation of index k is aimed for all individuals in the population.

To generate the offspring, each genotype was slightly modified.
A path in GRN was added, eliminated, or its sign was flipped with the probability $1/N^2$. Also, the parameter values $\gamma_i$, $D_i$, $\theta_i$ were modified by adding a random number from a Gaussian distribution $\eta(x)=\frac{1}{\sqrt{2\pi\sigma^2}}exp(-\frac{x^2}{2\sigma^2})$, while restricting these values to the set [0,1]. We set $\sigma=0.01$.

\subsubsection*{extracting working network}
Although the GRN is uniform for each cell in an individual, working gene expression dynamics differ from cell to cell. Additionally, the expression level of most genes changes discretely between epochs, indicating that the regulation of the output gene is different between epochs. Thus, it is difficult to describe gene expression regulation in one cell for a given epoch, because the whole GRN has too many edges, which may work for other cells during other epochs. To identify which regulations are essential at a certain location and time, we extracted part of the GRN that works at a certain location during a certain period by adopting the following systematic method.\\

1. We fixed a cell to analyze and trace the dynamics of the input to the output gene in that cell. The developmental time was divided into epochs by checking the changes in the input to the target, which are responsible for the epoch. In practice, we defined the time span during which the input goes out of the range $[-2/\beta:2/\beta]$, which is the dynamic range for the reaction term in the reaction diffusion equation (see equation (1) and (2)). Epochs are defined as the time span.\\

2. Fix the epoch to analyze and trace the dynamics of other genes during the epoch. If a gene is not expressed at all throughout the period, the regulation from and to the gene does not work. Hence, both input and output edges to such genes are eliminated.\\

3. Even if a gene is expressed for a certain time during the period, it does not necessarily mean that paths connecting to the gene are essential for the dynamics of the output gene’s expression. To check this point, we examine gene $j$'s expression as the input of another gene $i$ with $J_{ij}\neq 0$. We then, checked whether changes in gene $j$'s expression contributed to the expression of the gene $i$ as an input, while the latter's change stayed within the dynamic range $[-2/\beta:2/\beta]$. If this contribution, defined below, was larger than a certain threshold value, then the edge $j\rightarrow i$ is assumed to work. Otherwise the edge was eliminated.

The specific procedure is as follows: Consider an edge $j\rightarrow i$. For a given cell, let $\Delta x_k(t)$ as the expression change of gene $k$. 
 \[
 if\;\; |\displaystyle\sum_kJ_{ik}\Delta x_k(t)|\; > 0\;\; and\; -2/\beta < \displaystyle\sum_kJ_{ik}\Delta x_k(t) < 2/\beta,\quad\]\[ check\,\, whether\;\;  \frac{J_{ij}\Delta x_j(t)}{\displaystyle\sum_kJ_{ik}\Delta x_k(t)} > threshold
 \]
where $\displaystyle\sum_kJ_{ik}\Delta x_k(t)$ is the net input change of gene $i$. The threshold was 0.01. Delete all the edges $j\rightarrow i$ that do not satisfy the above condition.\\

4. Last, delete all the edges on genes that did not have a route from morphogens or to the output gene. 

\subsubsection*{overlap ratio}
The overlap between two networks $\bm{A}=\{ a_{i,j} \}\;\text{and}\; \bm{B}=\{ b_{i,j}\}$ was computed as follows.
\begin{equation}overlap= \frac{\sum{\theta(\bm{A},\bm{B},i,j)}}{min(\sum\phi(\bm{A},i,j), \sum\phi(\bm{B},i,j))}
\end{equation}

where
\[
\theta(\bm{A},\bm{B},i,j)
=\begin{cases}
1 & \text{if $a_{i,j}=b_{i,j}$}\\
0 &  \text{else}
\end{cases},\;\;\,
\phi(\bm{A},i,j)
=\begin{cases}
0 & \text{if $a_{i,j}=0$}\\
1 &  \text{else}
\end{cases}
\]
The numerator is the number of common edges between the two networks, while the denominator is the size of the smaller network. The overlap is 1 when a smaller network is completely included in the larger network and is 0 when the two networks do not have common edges. 

\section*{Acknowledgments}
 The authors would like to thank Tetsuhiro Hatakeyama, Shuji Ishihara,
and Naoki Irie for useful discussions. This work was partially
supported by a Grant-in-Aid for Scientific Research (No. 21120004)
on Innovative Areas "Neural creativity for communication" (No.
4103) and the Platform for Dynamic Approaches to Living Systems from
MEXT, Japan.

\renewcommand{\refname}{Literature cited}

\clearpage

\section*{Figure Legends}
\begin{center}
\includegraphics[width=6in]{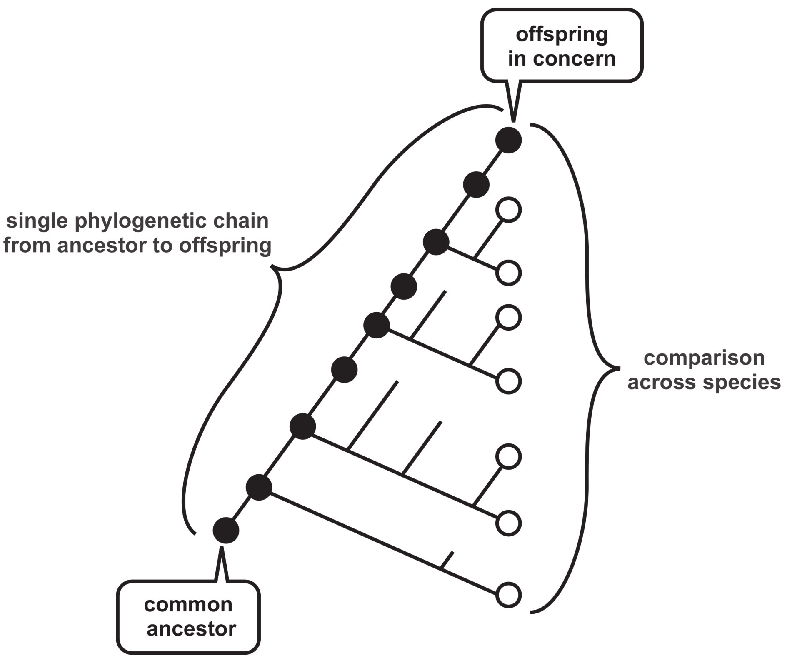}
\end{center}
\begin{quote}
\textbf{Figure 1.} \textbf{Schematic representation of single-chain-phylogeny and comparison across species in phylogenic tree}\newline
Schematic representation of the comparisons along  single-chain  phylogeny and across species. In the phylogenic tree shown here, the currently existing species represented by the right-end circles, are diverted from a common ancestor. Branching from a common ancestor leads to establishment of some new species, while some are terminated by extinction. The comparison of developmental processes across species is made over the existing species. On the other hand, a single phylogenic chain, which we focused in this study, is given as the line from the common ancestor to the offspring in concern. From the species in concern, ancestors are uniquely traced back. The comparison of developmental processes along this chain is possible at least in theory or simulations, which provides fundamental information on possible relationship between development and evolution.
\end{quote}
\clearpage

\begin{center}
\includegraphics[width=6in]{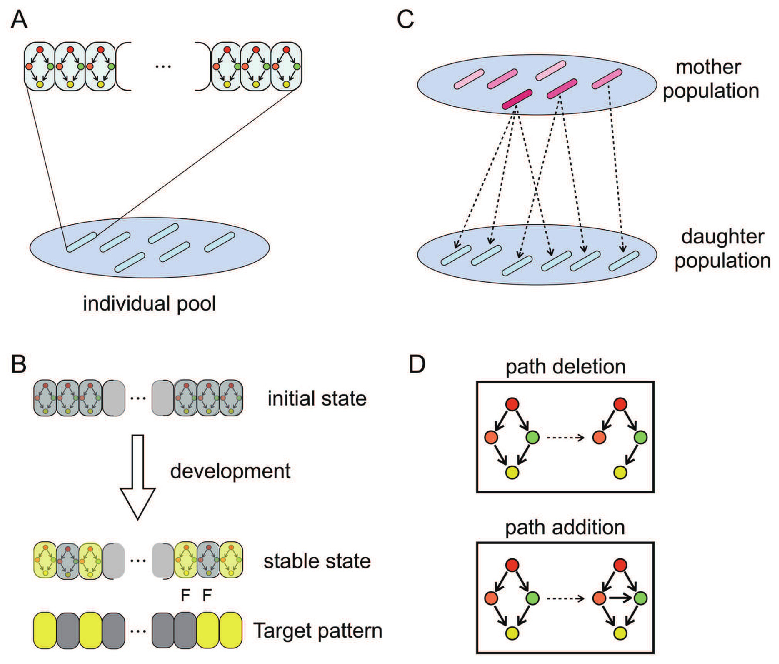}
\end{center}
\begin{quote}
\textbf{Figure 2. Schematic representation of simulation procedure.}\newline
(\textbf{A}): There are 100 individuals in a pool for each generation. Each individual consists of 96 uniform cells, which share a common GRN, while the GRN differs slightly between individuals.\newline
(\textbf{B}): Each individual develops from the same initial state in which genes are not expressed(i.e., with $x\sim0$) except for genes receiving the maternal gradient. Over time, individuals develop into stable states. Colors of cells indicate the expression level of the output genes; yellow is high, gray is low. After reaching a stable state, the expression pattern of the output gene was compared with the predefined target pattern. The fitness level was then elevated as the stable expression of the output gene approached the target pattern (see Methods for detail).\newline
(\textbf{C}): After the fitness of every individual was calculated, the population for the next generation was created. Each individual was selected as a mother with a probability proportional to its fitness. In the figure, the degree of red color indicates the fitness.\newline
(\textbf{D}): The GRN of a daughter is slightly different from the mother's, with a given mutation rate. The mutation involves deletion or addition of paths in the mother's GRN, and a change in characteristic parameters in expression dynamics and the diffusion constant (see Methods for details).
\end{quote}
\clearpage

\begin{center}
\includegraphics[width=6in]{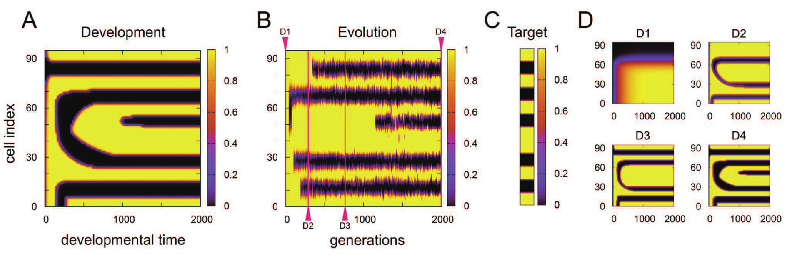}
\end{center}
\begin{quote}
\textbf{Figure 3. An example of space-time diagram of evolution and development.}\newline
(\textbf{A}): The expression level of the output gene is shown with developmental time as the horizontal axis and cell index (spatial position) as the vertical axis. The expression level of the output gene of the corresponding cell at a given time is color coded, (side bar) with black indicating the lowest and yellow indicating the highest expression levels. Development consists of a few epochs with rapid changes in the pattern, separated by quasi-stationary regimes with little change in the pattern, until the target pattern is shaped by development.\newline
(\textbf{B}): The spacetime diagram of the evolutionary course, corresponding to (\textbf{A}). The expression level of the final output gene (at time=2000) is shown with evolutionary generation as the horizontal axis and cell index (spatial position) as the vertical axis. This figure shows how the pattern is acquired through evolution. At each generation, the final pattern of the direct ancestor of the next generation is shown. The evolution of the developed output pattern consists of quasi-static regimes sandwiched by epochs with rapid change resulting from mutation, until the target pattern is evolved.\newline
(\textbf{C}): The predefined target pattern adopted in the present simulation.\newline
(\textbf{D}): Space-time diagram of the developmental process for several generations in (\textbf{B}). Each figure shows the development of the ancestral expression pattern at each generation, 0(D1), 300(D2), 750(D3), and 2000(D4). For reference, these generations are each marked by a red triangle at the top or bottom in (\textbf{B}).\newline
\end{quote}
\clearpage

\begin{center}
\includegraphics[width=4in]{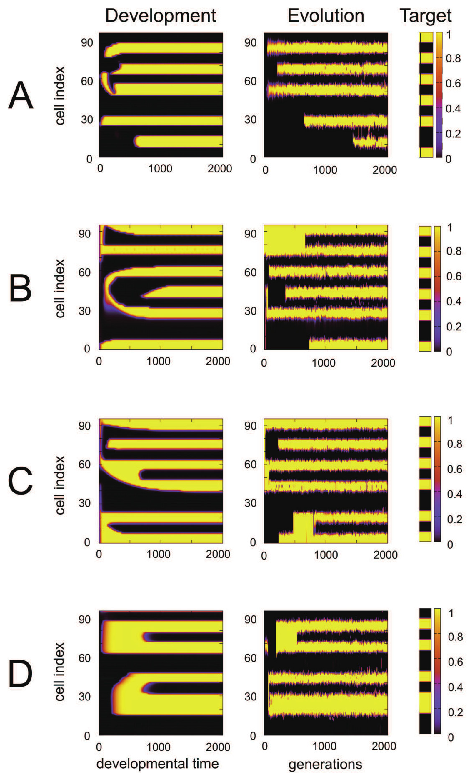}
\end{center}
\begin{quote}
\textbf{Figure 4. Four additional examples of evo-devo congruence.}\newline
(\textbf{A})-(\textbf{D}): Each row shows space-time diagrams of evolution and development, in the same way as Figure 3, although the target patterns are different. See supplemental information for additional examples.
\end{quote}
\clearpage

\begin{center}
\includegraphics[width=6in]{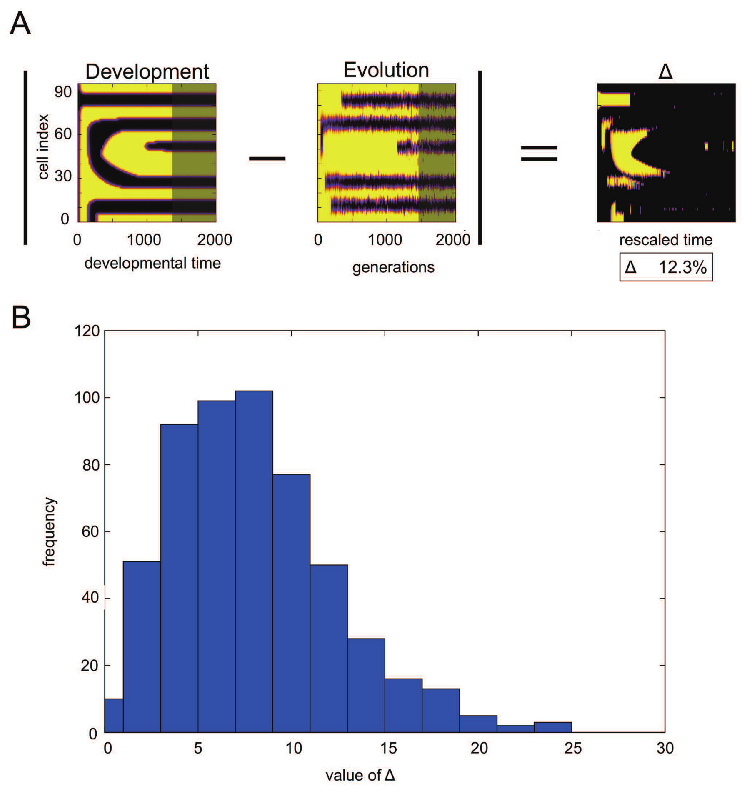}
\end{center}
\begin{quote}
\textbf{Figure 5. Quantitative analysis of the degree of evo-devo congruence.}\newline
(\textbf{A}): Schematic diagram illustrating quantitative analysis of the similarity between evolution and development. The differences between development and evolution were computed by subtracting expression levels at each pixel. By taking the absolute value of difference, and averaging the space-time pixels, the average difference was computed . To avoid over-estimating similarity, the region before the emergence of the first stripe and after the final pattern was ignored for both development and evolution. For example, the gray-masked region of the development and evolution figures does not include data for the calculation. If one stripe is completely shifted in time,  is approximately 8\%.\newline
(\textbf{B}): Histogram of the distribution of the $\Delta$ values. The abscissa is the $\Delta$ value computed via the procedure described in (\textbf{A}). The ordinate is the frequency of touch $\Delta$ values determined by bin size 2 . Distribution was obtained from 500 runs with different target patterns.
\end{quote}
\clearpage

\begin{center}
\includegraphics[width=6in]{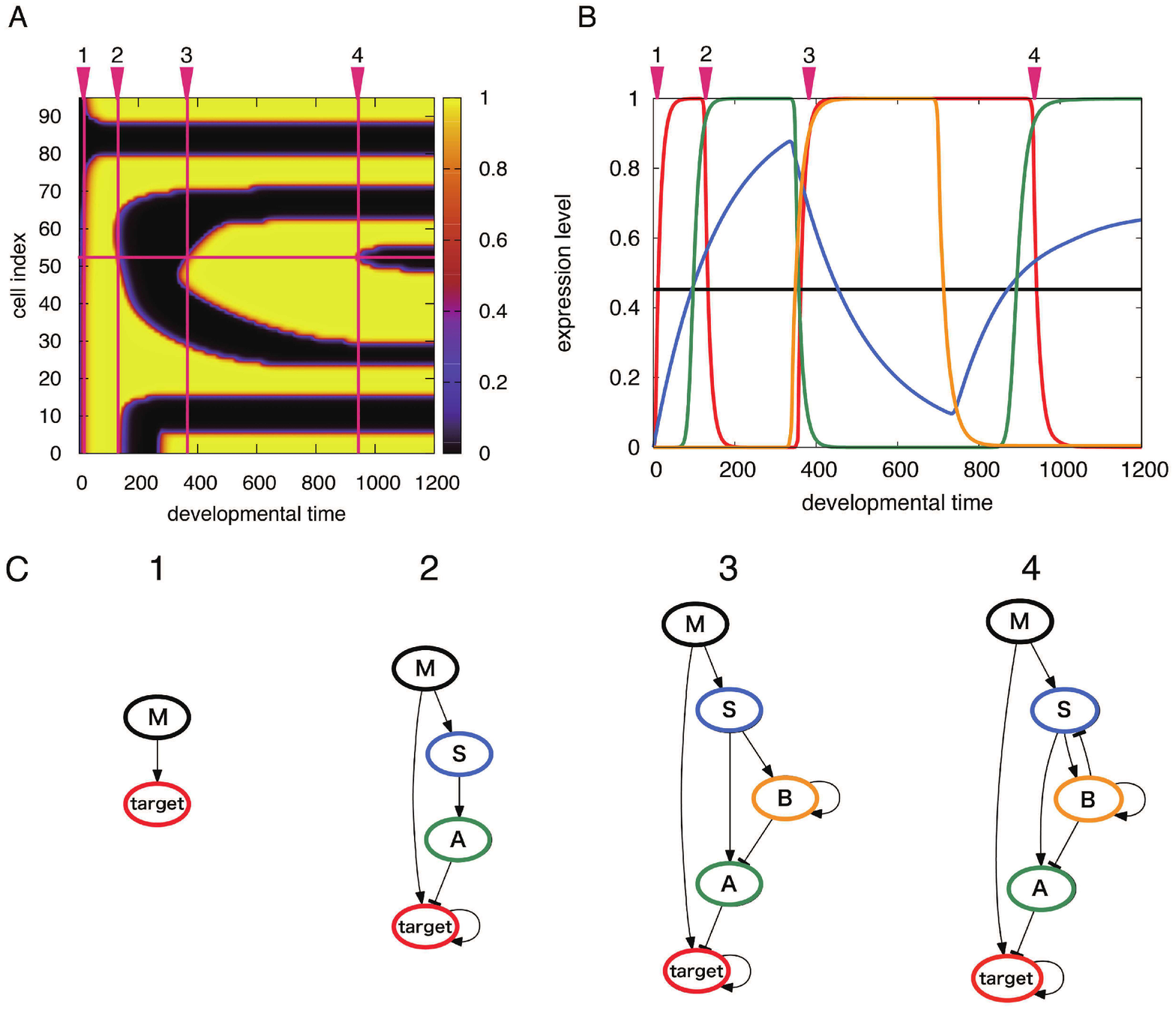}
\end{center}
\begin{quote}
\textbf{Figure 6. An example of developmental dynamics with several epochs for a single cell.}\newline
(\textbf{A}): Developmental space-time diagram of the output expression level. The developmental time up to 1200 in Figure 3 is zoomed, in order to clearly distinguish epochs. There are four epochs associated with one cell, marked by crossing of the red lines.\newline
(\textbf{B}): Gene expression dynamics at the cell highlighted by red horizontal line in (\textbf{A}). The time series of the expression level is plotted where the line color corresponds to that of network node in (\textbf{C}), which representing each gene. Most genes change their expression level within each epoch, except for the gene S represented by the blue line.\newline
(\textbf{C}): The working networks that function to switch the output expression at each epoch. These networks are derived from analyses on gene expression dynamics(see method). Arrow edges indicate positive regulation while the headed edges indicate negative regulation.
\end{quote}
\clearpage

\begin{center}
\includegraphics[width=4in]{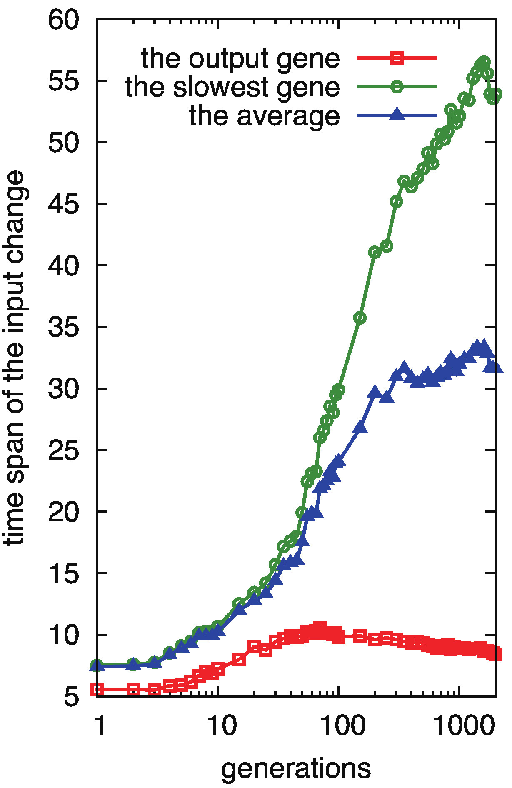}
\end{center}
\begin{quote}
\textbf{Figure 7. Evolution of the time scale for the input for the output and other genes}\newline
By taking genes whose expression level change between on and off for each developmental epoch, the time scales are computed as the time span that the input for the gene passes through the dynamic range during each development. The red square gives the time span for the input of the output gene, and the blue triangle(green circle) denotes the average (the largest) of the time span among the genes that have a path to the output gene, respectively. The time spans are computed from the average of 500 samples of evolution simulations.
\end{quote}
\clearpage

\begin{center}
\includegraphics[width=6in]{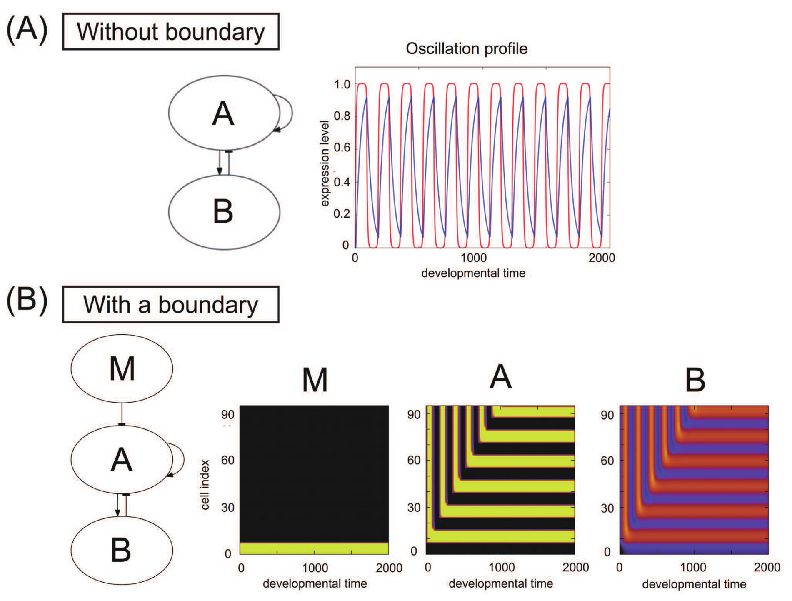}
\end{center}
\begin{quote}
\textbf{Figure 8. Feedback oscillation and its fixation.}\newline
(\textbf{A})\textbf{Without boundary:} Minimal network for oscillatory expression with the time series of the expression for a specific cell. Gene A activates the expression of gene B and itself, and gene B suppresses A. In the plotted time series, developmental time is plotted as the abscissa, and the expression levels of A (red) and B (blue) are plotted as the ordinate.\newline
(\textbf{B})\textbf{With a boundary:} The input from gene M, which was influenced by the maternal factor, was included in the oscillatory network. The space-time diagram of genes M, A, and B illustrate how oscillatory expressions of gene A and gene B were fixed to form stripes. Gene M was expressed near the boundary.\newline
\end{quote}
\clearpage

\begin{center}
\includegraphics[width=4in]{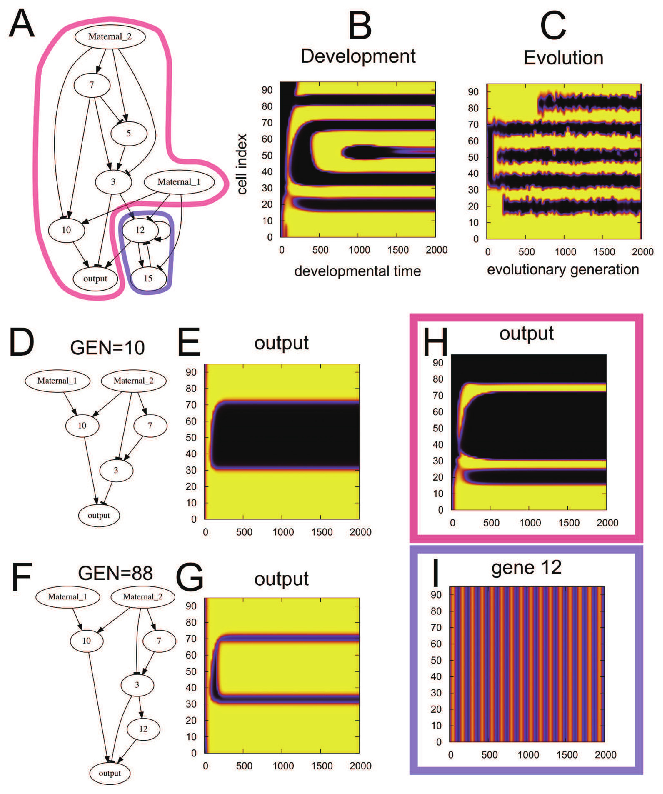}
\end{center}
\begin{quote}
\textbf{Figure 9. Example of evo-devo congruence with network structures}\newline
(\textbf{A}): An example of a core part of the GRN at the 2000th generation, evolved to achieve the target pattern. From the maternal factors, the feedforward networks is surrounded by magenta, while the network module for feedback oscillation, consisting of genes 12 and 15, is surrounded by blue. Here, genes and paths that are not essential to the output pattern formation were eliminated.\newline
(\textbf{B,C}):Space-time diagrams of the output gene expression for development (\textbf{B}) and evolution (\textbf{C}) or the GRN are displayed together to show the degree of similarity between them. The vertical axis denotes the space (cell index), and the horizontal axis denotes either evolutionary generation (evolution) or developmental time (development). For this example, the $\Delta$ value is 8.0\%.\newline
(\textbf{D}): An example of the core part of the GRN at the 10th generation (i.e., very early generation) and the corresponding space-time expression diagrams of the output.\newline
(\textbf{E}). Feedforward structure of the GRN is evolved at this early stage of evolution. The vertical axis of the phase diagram denotes the space (cell index), and the horizontal axis denotes developmental time. This expression is observed at a very early stage of development in (\textbf{B}), at approximately the 10th generation.\newline
(\textbf{F}): The network structure at the 88th generation. Through evolution, feedforward structures are sequentially acquired in the downstream region of the core part of the GRN.\newline
(\textbf{G}): Developmental space-time diagram of the expression of the output gene for the network \textbf{F}. This expression profile provides the top and bottom stripes in (\textbf{B}).\newline
(\textbf{H}): Developmental space-time diagram of the output expression of the 2000th generation where the feedback oscillation module is eliminated. Without feedback, only part of (\textbf{B}) is generated.\newline
(\textbf{I}):Developmental space-time diagram of the expression of gene 12, one of the feedback modules in (\textbf{A}), which produces a spatially homogeneous and temporally periodic oscillation if constant activation is applied by gene 3. The combination of this feedback oscillation and the boundary condition provided by gene 12 shown here produces the three internal stripes in (\textbf{B}).
\end{quote}
\clearpage

\begin{center}
\includegraphics[width=6in]{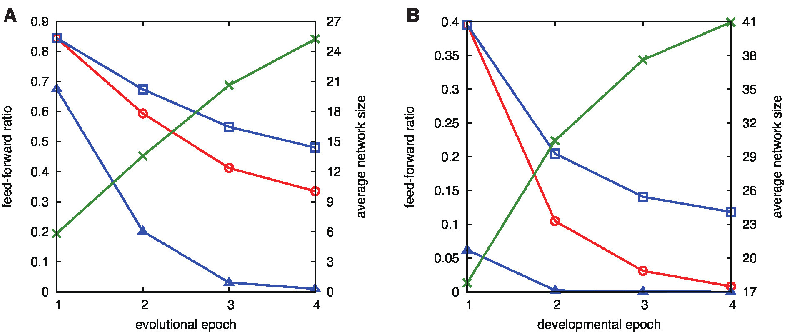}
\end{center}
\begin{quote}
\textbf{Figure 10. Feedforward ratio}\newline
(\textbf{a}):The red circle denotes the fraction of working networks that do not include a feedback loop, plotted as a function of evolutional epoch. The fraction is computed from 500 samples of evolution simulations. For reference, the probability estimated from the value at the first epoch only as a result of the increase in the network size is plotted as the blue square, while the triangle denoted such probability computed from $2\times10^6$ random networks of the corresponding size, generated only under the constraint that there are paths from the morphogens, and to the target.The average size of the network computed from the simulation is also plotted as cross with the second vertical axis.\newline
(\textbf{b}):Ratio of the feedforward network as a function of each developmental epoch, computed in the same manner as (\textbf{a}) and the same use of symbols.\newline
\end{quote}
\clearpage

\begin{center}
\includegraphics[width=6in]{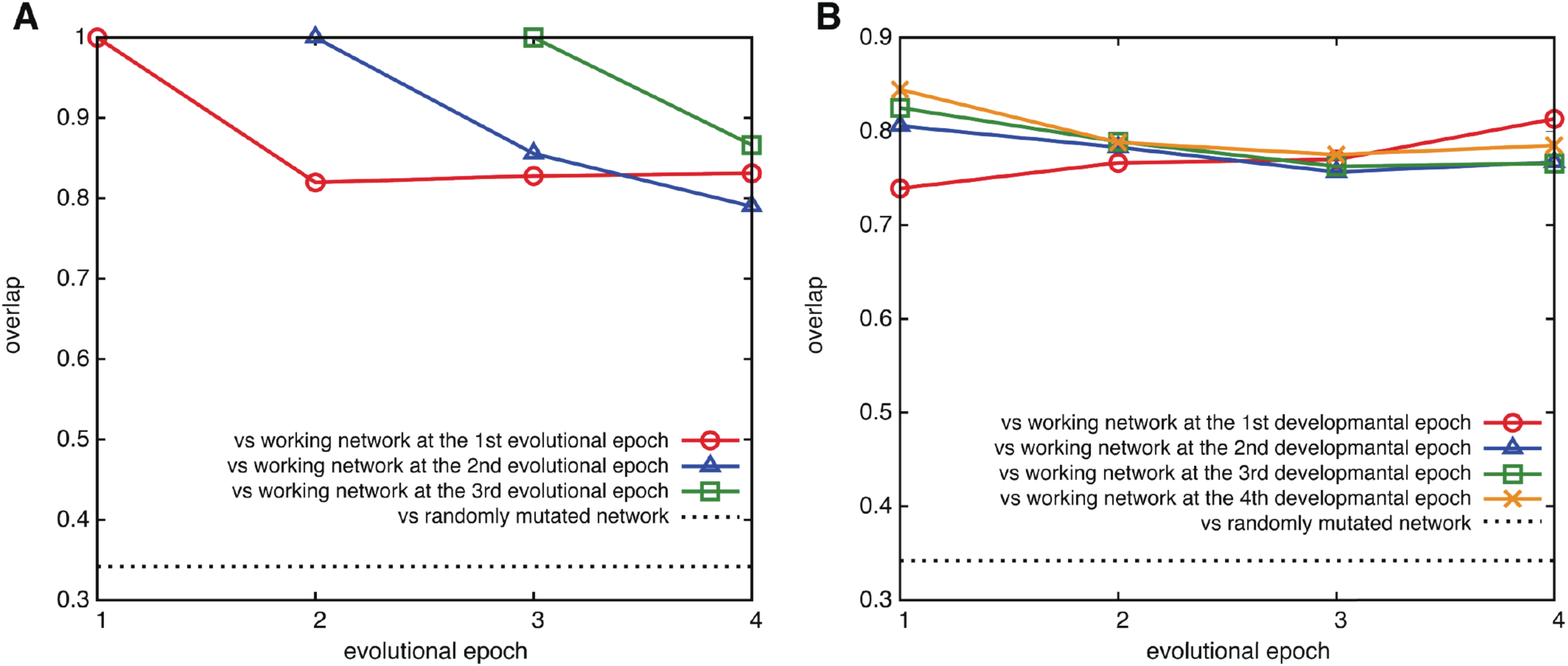}
\end{center}
\begin{quote}
\textbf{Figure 11. Network overlap}\newline
(\textbf{A})The overlap of the working network at each epoch with its ancestral network. The overlap of the 1st(\protect\raisebox{1pt}{{\footnotesize $\bigcirc$}}), the 2nd({\small $\triangle$}) and the 3rd({\normalsize $\square$}) epochs with the offspring network at later epochs are plotted. See method for the definition of the overlap, which is computed as a statistical average from 500 samples of evolution. Dashed line is the average overlap between randomly generated network and the same network that underwent 2000 sttif of random mutations.\newline
(\textbf{B})The overlap of the working networks at the 1st(\protect\raisebox{1pt}{{\footnotesize $\bigcirc$}}), the 2nd({\small $\triangle$}), the 3rd({\normalsize $\square$}) and the 4th(\protect\raisebox{1pt}{{\normalsize $\times$}}) \textit{developmental} epoch with the ancestral networks at each \textit{evolutional} epoch represented by the abscissa. Dashed line is the average overlap between randomly generated network and the same network that unnderwent 2000 sttif of random mutations.\newline
\end{quote}
\clearpage

\begin{center}
\includegraphics[width=6in]{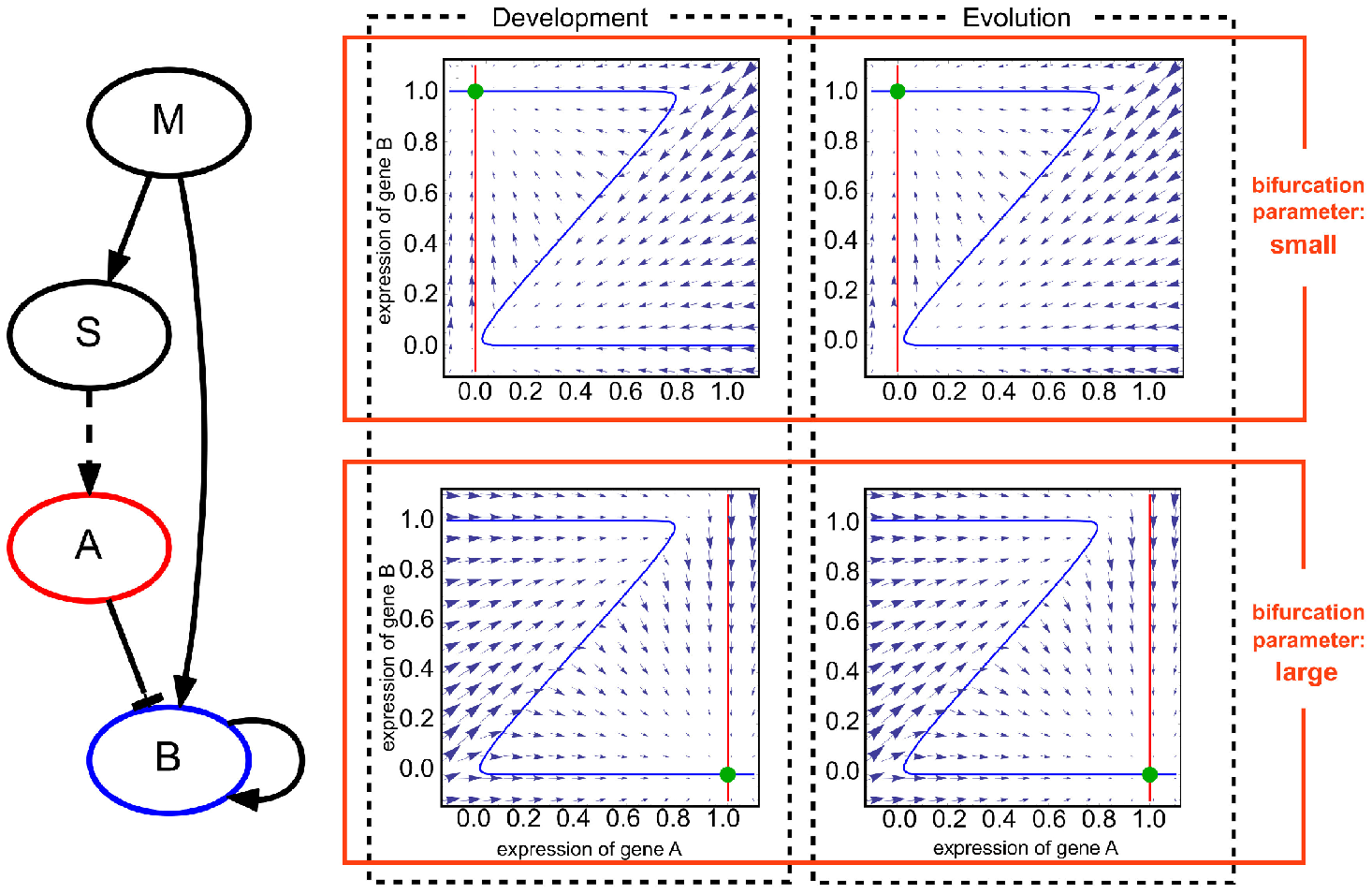}
\end{center}
\begin{quote}
\textbf{Figure 12. Evolution and development as bifurcation.}\newline
The network structure where the expression level of gene S changes slowly (left).
Phase space diagrams plotting the expression levels of gene A (horizontal access) and gene B (vertical access). The blue line represents the nullcline of gene A, and the red line represents the nullcline of gene B (right). The green circle denotes the final stable cell state from the initial conditions in each of the diagrams.
\textbf{Development(left column):} Expression level of the slow variable works as a bifurcation parameter. While the expression of gene S is lower than the threshold of gene A, the stable fixed point can be found at approximately ($x_A\sim0$,$x_B\sim1$)(upper left). As development progresses, the expression level of gene S increases, and after the expression of gene 1 exceeds the threshold of gene A, the nullcline of gene A shifts slightly to the right, indicating a higher value (lower left). Gene A inhibits the expression of gene B, so that the fixed point is changed to ($x_A\sim$1,$x_B\sim0$).\newline
\textbf{Evolution(right column):}
Phase diagram representing the expression levels of gene A (horizontal axis) and gene B (vertical axis). The blue line represents the nullcline of gene A and the red line represents the nullcline of gene B. The activation strength from gene A to gene B is regarded as a continuous value here. If the activation strength is low, the expression level of gene 1 is low, so that the stable fixed point is observed at approximately ($x_A\sim0,x_B\sim1$)(upper right). When the strength is sufficiently large, the expression level of gene 1 assumes a higher value so that the fixed point is observed at approximately ($x_A\sim1,x_B\sim1$)(lower right).
 Note that by comparing these two columns, a strong correspondence is observed in bifurcation between evolution and development.
\end{quote}
\clearpage

\begin{center}
\includegraphics[width=4.5in]{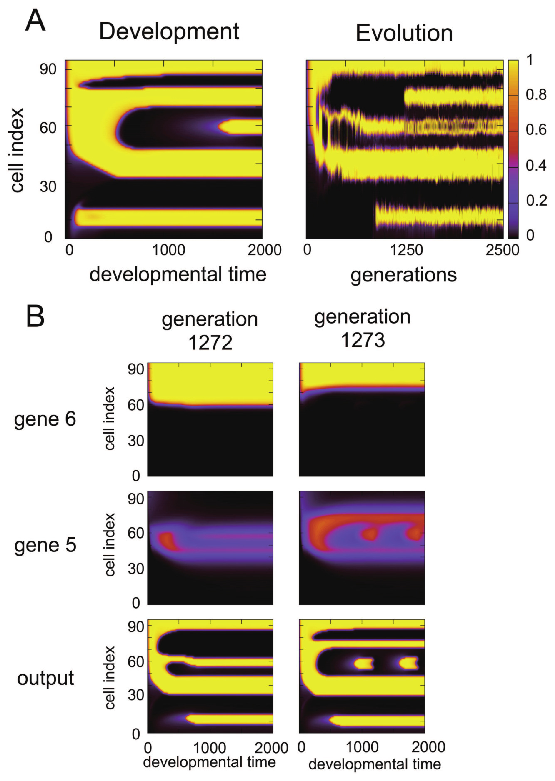}
\end{center}
\begin{quote}
\textbf{Figure 13. Violation of evo-devo congruence.}\newline
(\textbf{A}): \textbf{Evolution:} The expression level of the final output gene (at time=2000) is shown with the generation (horizontal axis) and cell index (vertical axis). The color scale is presented as a side bar, as in Figure 3B. According to the figure, the second upper stripe is acquired at the most recent stage of evolution, and the first, third and fourth upper stripes branch from the same root, so that the second stripe emerges from the first upper valley.\newline
\textbf{Development:} Space-time diagram of the expression with developmental time (horizontal axis) and cell index (vertical axis). The third upper stripe emerges at the most recent stage of development. Unlike evolution, the first, second and fourth upper stripes branch from the same root, and the third stripe emerges from the second upper valley. Here, evo-devo congruence is topologically violated.\newline
(\textbf{B}): Developmental diagrams plotted for generations 1272 and 1273. These genes show drastic change in their expression between the two generations. Gene 6 provides a feedforward regulation to the output, and inhibits the expression of gene 5, which is a component of the feedback loop to generate oscillatory expression. A mutation, which adds a path to gene 6, occurs between the two generations, which inhibits the expression of gene 6. Through this mutation, the expression of gene 6 is suppressed, thus shrinking the resulting stripe, and producing an additional stripe for gene 5. With this change, the ordering of the expression of the output gene is altered.
\end{quote}
\clearpage
\clearpage
\section*{Supplementary Material}

\begin{center}
\includegraphics[width=6in]{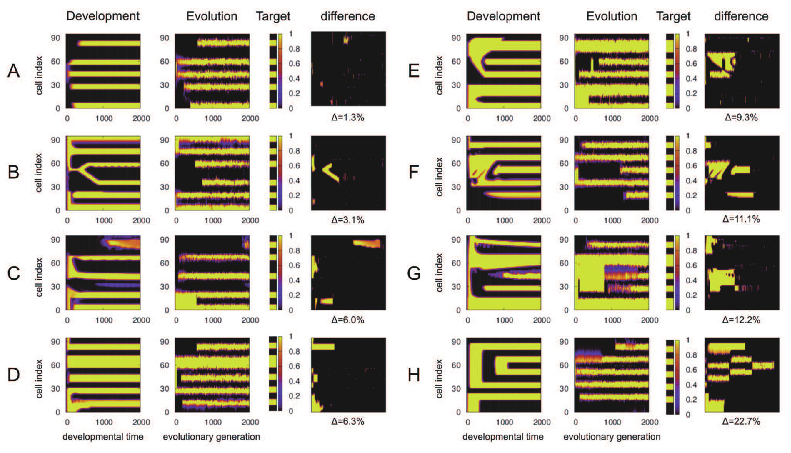}
\end{center}
\begin{quote}
\textbf{Figure S1.} \textbf{Space-time diagrams of evolution, development and differences between the two.}\newline
(\textbf{A})-(\textbf{H}) Eight additional space-time comparisons between evolution and development. Each consists of a space-time diagram of development, evolution, preset target pattern and difference between evolution and development. Space-time diagrams of evolution and development and their target pattern are plotted in the same way as in Figs.3 and 4. Calculated values of the difference $\Delta$ are shown below the diagram. For A-D, the evolution and development corresponded well, while a clear violation was observed for H.
\end{quote}
\clearpage

\begin{center}
\includegraphics[width=6in]{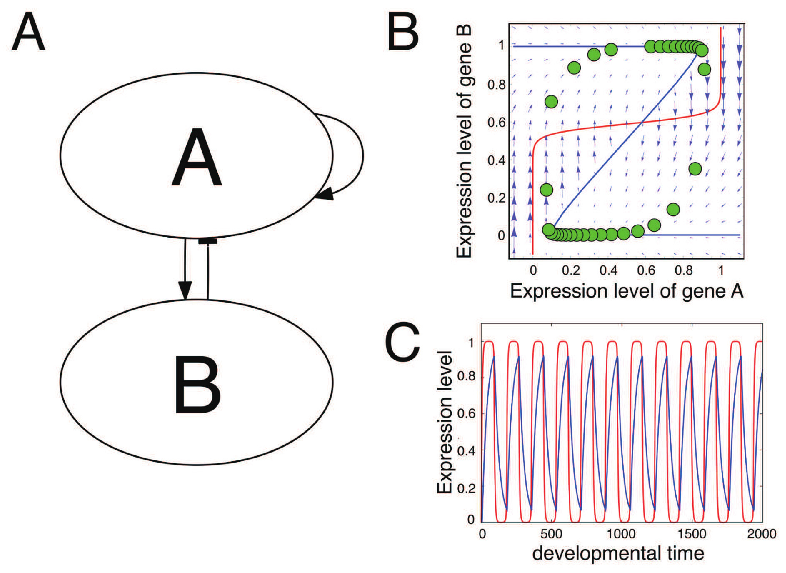}
\end{center}

\begin{quote}\textbf{Figure S2} \textbf{Detailed analysis on feedback oscillation.}\newline
(\textbf{A}): network structure for data presented  in Figure 8.\newline
(\textbf{B}): Phase diagram of the expression dynamics. Two nullclines of gene expression cross at a single, unstable fixed point, and the cell state will oscillate on a limited cycle. Green circles represent the cell state at time step intervals of 5, within a single cycle. The distance between two nullclines is shortest at the upper right and lower left corners so that cell state changes are slower at these corners.\newline
(\textbf{C}): Time profile of the feedback oscillation for a specific cell. The abscissa represents developmental time and the ordinate is expression level. Gene A is plotted as a red line, while gene B plotted as a Blue line.
\end{quote}
\clearpage

\begin{center}
\includegraphics[width=5.4in]{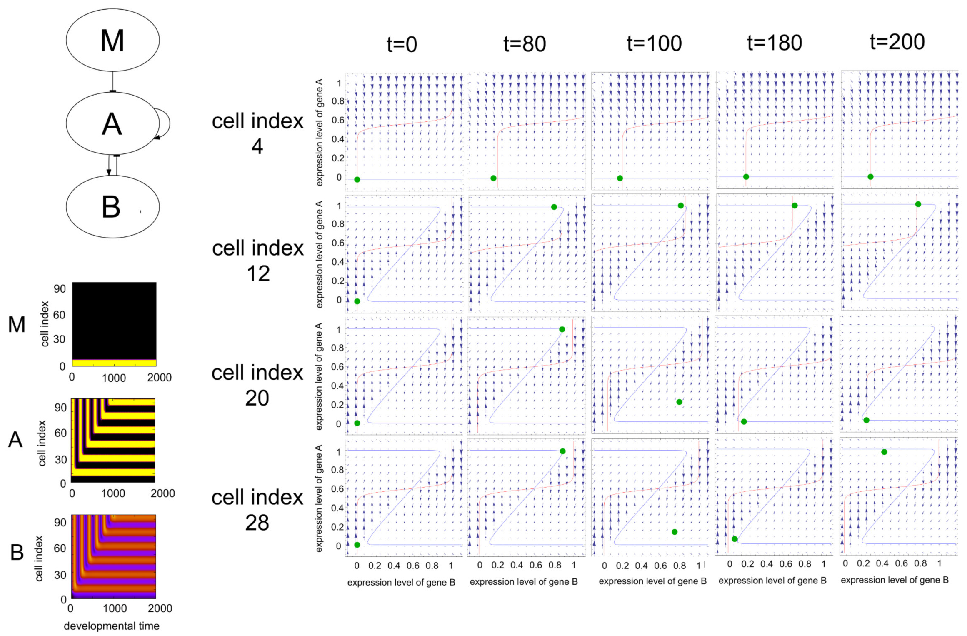}
\end{center}
\begin{quote}
\textbf{Figure S3} \textbf{Temporal change of flow in the phase space over cells.}\newline
The oscillation fixation mechanism is revealed through comparison of flow temporal changes in the phase space over cells. The central cells of the first two stripes (cell indices 12 and 28) and valleys (cell indices 4 and 20).\newline
\newline
(1)\textbf{t=0}:\newline
 Both gene A and gene B assume a null value, and gene B is inhibited by gene M within the first 8 cells. With these two initial conditions, flow in the phase space where cell index = 4 is different from other cells. At a stable fixed point therein, both the expressions of gene A and gene B are low (i.e., a low-low state ). This fixed point is the root of the first valley.\newline
(2)\textbf{t=80}:\newline
 As development begins, expression of non-inhibited cells begin to oscillate and move towards a state where both expressions of gene A and gene B are high (i.e., a high-high state), while cells of the first valley maintain the slow-low state. Thus, protein A diffuses from the first stripe to the first valley. Due to the incoming diffusion of protein expression of gene A, at cell index 4, the nullcline of gene A slides to the right, so that the expression of gene A assumes a higher value. Correspondingly, at cell index 8, the nullcline of gene A slides to the left, and crosses the nullcline of gene B to create a fixed point.\newline
(3)\textbf{t=100}:\newline
 The expression of cell index 4 is constrained  to the newly formed high-high fixed point. However, the expressions of cells at index 20 and 28 continued to oscillate.\newline
(4)\textbf{t=180}:\newline
 Through the oscillation, cell index 20 approaches a low-low state for the second time, and at this time, protein B at the first stripe diffuses to the second valley. Thus, nullclines slide in the cells at indices 12 and 20 to the left and right, respectively. The nullcline of gene B then crosses, at a low-low state, in the cell at index 20.\newline
(5)\textbf{t=200}:\newline
 The expression level at cell index 28 continued to oscillate while that at cell index 20 was constrained at the newly formed low-low state fixed point, similar to cell index 12 at t=80. Protein B subsequently diffused from the cell index 28 to the second valley, which resulted in the emergence of the second stripe. In this way stripes were shaped from the oscillation.
\end{quote}
\clearpage

\textbf{Text S1} \textbf{Mathematical analysis of the slow variable}

Here we considered the relaxation time around a fixed point. For simplicity we considered the case with $\gamma_i=1$, to demonstrate that the relaxation time is longer even without the change in $\gamma$.
The stability of the fixed point was given by eigenvalues of the Jacobian matrix $W_{ij}$ where the diagonal component $W_{ii}$ is given by $-1$ and the off-diagonal component $W_{ij}$ is given by $J_{ij} \beta exp(-\beta X_i)/(1+exp(-\beta X_i))^2$ where $X_i=\sum_j J_{ij}(x_j-\theta_j)$. If $x_j$'s are close to 0 or 1, their deviation from $\theta_j$ is sufficiently larger than the detection threshold $1/\beta$, the off-diagonal elements are close to zero, and the eigenvalues are given by -1 (or -$\gamma_i$ if it is not 1). When $x_j$'s takes on intermediate values closer to $\theta_i$, the off-diagonal elements assume larger values, and the eigenvalues are shifted from -1, either upwards or downwards. Hence, some exponents approach zero. As long as the real components of the eigenvalues are negative, the fixed point remains stable, but the stability is weaker, with the exponent closer to zero. This results in an increase in the timescale of the relaxation, given by the inverse of the real component of the eigenvalue. With this mechanism, the slowly changing variable is generated even without small $\gamma_i$.
\clearpage

\begin{center}
\includegraphics[width=6in]{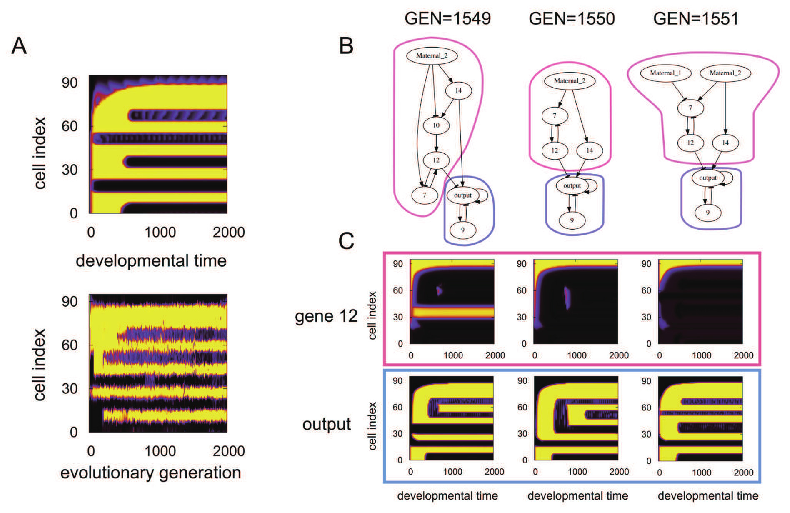}
\end{center}
\begin{quote}
\textbf{Figure S4} \textbf{Network analysis of the extra example of the violation of evo-devo congruence 1.}\newline
 In development, the third and fourth upper stripes stem from the some root, while in evolution the top three stems originate from the same root. This branching change occurs during generation 1549 where the second and third branches are clearly stabilized. In this case, unlike the former examples, topological changes in branching occur sequentially during three generations. A time-space diagram of the output and gene 7 are presented in Figure B. Gene 7 exhibits two stripes from generation 1549, which are driven by a feedforward mechanism. Due to the boundary effect of gene 7, the upper three stripes are generated in the output gene.\\

Then, during generation 1550, part of the feedforward mechanism upstream of gene 7 is deactivated, which enhances the region expressed by the feedback oscillation mechanism. As a result, the upper four stripes that emerge share the same oscillation mechanism.\\

At generation 1551, mutation occurs upstream of gene 7, so that the morphogen comes to inhibit the remaining feedforward mechanism. Before the mutation, gene 7 exhibits weak temporal expression in cell sites 30-85. After the mutation, this temporal expression is inhibited so that the expression region is restricted to cell sites 60-85. Due to this change, the third and the fourth upper stripes emerge faster than the first and second stripes, while the third stripe, generated in advance of the first two, provided a boundary to generate the second stripe.\\

To summarize, the upper 4 stripes were generated by the feedback oscillation mechanism, but the change in the boundary condition due to mutation in the upstream feedforward mechanism introduced the branching combination.
\end{quote}
\clearpage

\begin{center}
\includegraphics[width=5.2in]{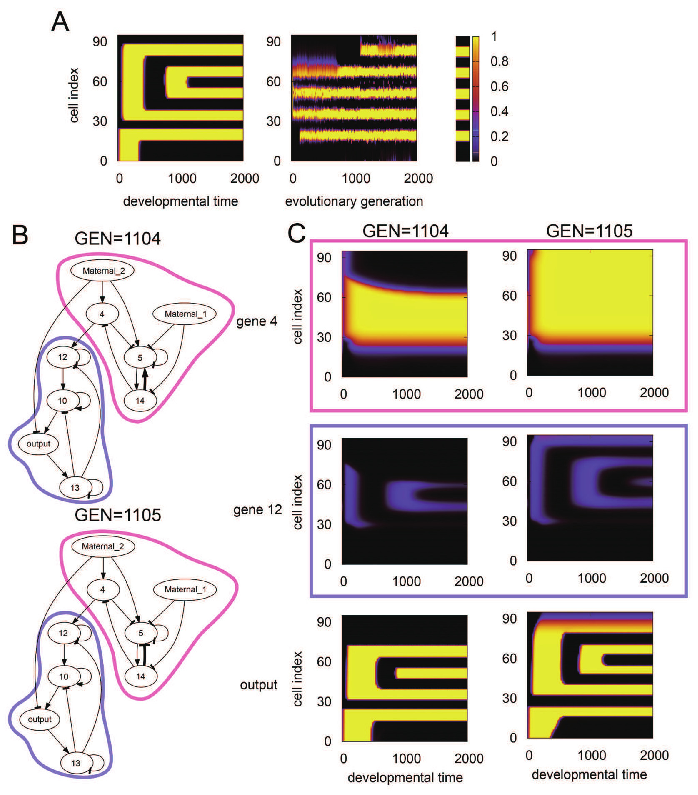}
\end{center}
\begin{quote}
\textbf{Figure S5} \textbf{Network analysis of an additional example of the violation of evo-devo congruence.}\newline
In evolution, the three central stripes are acquired nearly simultaneously, and two additional stripes are subsequently acquired independently. However, in development, at the final evolved generation, the 1st and 4th stripes were generated from the same root at the same time, and subsequently the 2nd and 3rd stripes were generated from a common root. These two branchings follow the oscillation-fixation mechanism. Only the bottom stripe is generated independently. The developmental order of stripe formation was acquired, between generations 1104 and 1105.\\

Genes that exhibited relevant change are displayed in Figure B. The expression of an upstream gene (green in the GRN figure below) and the downstream gene (red in the GRN figure) are plotted at the upper and lower columns, respectively. In this example, the feedforward mechanism worked only temporally, as shown in the transient expressed before time step = 100 (Figure C). This temporal expression region also corresponded to the region of feedback oscillation. Due to the mutation, the spatial domain of the transient expression was extended upward. Violation of evo-devo congruence was therefore induced by this expansion of the transient expression region.
\end{quote}


\begin{thebibliography}{49}
\providecommand{\natexlab}[1]{#1}
\expandafter\ifx\csname urlstyle\endcsname\relax
  \providecommand{\doi}[1]{doi:\discretionary{}{}{}#1}\else
  \providecommand{\doi}{doi:\discretionary{}{}{}\begingroup
  \urlstyle{rm}\Url}\fi

\bibitem[{Alon(2006)}]{alon2006introduction}
Alon U. 2006.
\newblock An Introduction to Systems Biology: Design Principles of Biological
  Circuits.
\newblock Chapman \& Hall/CRC Mathematical and Computational Biology. London:
  Taylor \& Francis.

\bibitem[{Ancel and Fontana(2000)}]{ancel2000plasticity}
Ancel LW, Fontana W. 2000.
\newblock Plasticity, evolvability, and modularity in {RNA}.
\newblock J Exp Zool B 288:242--283.

\bibitem[{Carroll et~al.(2009)Carroll, Grenier and Weatherbee}]{carroll2009dna}
Carroll SB, Grenier J, Weatherbee S. 2009.
\newblock From DNA to diversity: molecular genetics and the evolution of animal
  design.
\newblock Hoboken, NJ: Wiley.

\bibitem[{Chance et~al.(1967)Chance, Pye and Higgins}]{chance1967waveform}
Chance B, Pye K, Higgins J. 1967.
\newblock Waveform generation by enzymatic oscillators.
\newblock Spectrum, IEEE 4:79--86.

\bibitem[{Ciliberti et~al.(2007)Ciliberti, Martin and
  Wagner}]{ciliberti2007robustness}
Ciliberti S, Martin OC, Wagner A. 2007.
\newblock Robustness can evolve gradually in complex regulatory gene networks
  with varying topology.
\newblock PLoS Computational Biology 3:e15.

\bibitem[{Cooke and Zeeman(1976)}]{cooke1976clock}
Cooke J, Zeeman E. 1976.
\newblock A clock and wavefront model for control of the number of repeated
  structures during animal morphogenesis.
\newblock Journal of Theoretical Biology 58:455--476.

\bibitem[{Cotterell and Sharpe(2010)}]{cotterell2010atlas}
Cotterell J, Sharpe J. 2010.
\newblock An atlas of gene regulatory networks reveals multiple three-gene
  mechanisms for interpreting morphogen gradients.
\newblock Mol Syst Biol 6:425.

\bibitem[{Darwin(1859)}]{darwin1859origin}
Darwin C. 1859.
\newblock On the Origin of Species by Means of Natural Selection, Or, The
  Preservation of Favoured Races in the Struggle for Life.
\newblock London: John Murray.

\bibitem[{De~Wit et~al.(1996)De~Wit, Lima, Dewel and
  Borckmans}]{PhysRevE.54.261}
De~Wit A, Lima D, Dewel G, Borckmans P. 1996.
\newblock Spatiotemporal dynamics near a codimension-two point.
\newblock Phys Rev E 54:261--271.
\newblock \doi{10.1103/PhysRevE.54.261}.

\bibitem[{Domazet-Lo{\v s}o and Tautz(2010)}]{Domazet-Loso:2010cs}
Domazet-Lo{\v s}o T, Tautz D. 2010.
\newblock A phylogenetically based transcriptome age index mirrors ontogenetic
  divergence patterns.
\newblock Nature 468:815--8.
\newblock \doi{10.1038/nature09632}.

\bibitem[{Eldredge and Gould(1972)}]{eldredge1972punctuated}
Eldredge N, Gould SJ. 1972.
\newblock Punctuated equilibria: an alternative to phyletic gradualism.
\newblock Models in paleobiology 82:115.

\bibitem[{Fran{\c c}ois et~al.(2007)Fran{\c c}ois, Hakim and
  Siggia}]{Francois:2007rm}
Fran{\c c}ois P, Hakim V, Siggia ED. 2007.
\newblock Deriving structure from evolution: metazoan segmentation.
\newblock Mol Syst Biol 3:154.
\newblock \doi{10.1038/msb4100192}.

\bibitem[{Fran{\c c}ois and Siggia(2012)}]{Francois:2012yg}
Fran{\c c}ois P, Siggia ED. 2012.
\newblock Phenotypic models of evolution and development: geometry as destiny.
\newblock Curr Opin Genet Dev 22:627--33.
\newblock \doi{10.1016/j.gde.2012.09.001}.

\bibitem[{Freitas et~al.(2012)Freitas, G{\'o}mez-Mar{\'\i}n, Wilson, Casares
  and G{\'o}mez-Skarmeta}]{freitas2012hoxd13}
Freitas R, G{\'o}mez-Mar{\'\i}n C, Wilson JM, Casares F, G{\'o}mez-Skarmeta JL.
  2012.
\newblock \textit{Hoxd13} Contribution to the Evolution of Vertebrate
  Appendages.
\newblock Developmental cell 23:1219--1229.

\bibitem[{Fujimoto et~al.(2008)Fujimoto, Ishihara and Kaneko}]{Fujimoto:2008jk}
Fujimoto K, Ishihara S, Kaneko K. 2008.
\newblock Network evolution of body plans.
\newblock PLoS One 3:e2772.
\newblock \doi{10.1371/journal.pone.0002772}.

\bibitem[{Glass and Kauffman(1973)}]{glass1973logical}
Glass L, Kauffman SA. 1973.
\newblock The logical analysis of continuous, non-linear biochemical control
  networks.
\newblock Journal of Theoretical Biology 39:103--129.

\bibitem[{Goodwin(1963)}]{goodwin1963temporal}
Goodwin BC. 1963.
\newblock Temporal organization in cells: A dynamic theory of cellular control
  processes.
\newblock London and New York: Academic Press.

\bibitem[{Goto and Kaneko(2013)}]{PhysRevE.88.032718}
Goto Y, Kaneko K. 2013.
\newblock Minimal model for stem-cell differentiation.
\newblock Phys Rev E 88:032718.
\newblock \doi{10.1103/PhysRevE.88.032718}.

\bibitem[{Gould(1977)}]{gould1977ontogeny}
Gould SJ. 1977.
\newblock Ontogeny and phylogeny.
\newblock Cambridge, MA: Harvard University Press.

\bibitem[{Hall(1999)}]{Hall1999evolutionary}
Hall BK. 1999.
\newblock Evolutionary developmental biology.
\newblock Berlin: Springer.

\bibitem[{Hall(2000)}]{hall2000balfour}
Hall BK. 2000.
\newblock Balfour, {G}arstang and de {B}eer: the first century of evolutionary
  embryology.
\newblock American Zoologist 40:718--728.

\bibitem[{Hazkani-Covo et~al.(2005)Hazkani-Covo, Wool and
  Graur}]{Hazkani-Covo:2005pb}
Hazkani-Covo E, Wool D, Graur D. 2005.
\newblock In search of the vertebrate phylotypic stage: a molecular examination
  of the developmental hourglass model and von {B}aer's third law.
\newblock J Exp Zool B 304:150--8.
\newblock \doi{10.1002/jez.b.21033}.

\bibitem[{Horikawa et~al.(2006)Horikawa, Ishimatsu, Yoshimoto, Kondo and
  Takeda}]{horikawa2006noise}
Horikawa K, Ishimatsu K, Yoshimoto E, Kondo S, Takeda H. 2006.
\newblock Noise-resistant and synchronized oscillation of the segmentation
  clock.
\newblock Nature 441:719--723.

\bibitem[{Irie and Kuratani(2011)}]{Irie:2011gd}
Irie N, Kuratani S. 2011.
\newblock Comparative transcriptome analysis reveals vertebrate phylotypic
  period during organogenesis.
\newblock Nat Commun 2:248.
\newblock \doi{10.1038/ncomms1248}.

\bibitem[{Ishihara et~al.(2005)Ishihara, Fujimoto and
  Shibata}]{ishihara2005cross}
Ishihara S, Fujimoto K, Shibata T. 2005.
\newblock Cross talking of network motifs in gene regulation that generates
  temporal pulses and spatial stripes.
\newblock Genes to Cells 10:1025--1038.

\bibitem[{Jaeger et~al.(2012)Jaeger, Irons and Monk}]{jaeger2012inheritance}
Jaeger J, Irons D, Monk N. 2012.
\newblock The inheritance of process: a dynamical systems approach.
\newblock Journal of Experimental Zoology Part B: Molecular and Developmental
  Evolution 318:591--612.

\bibitem[{Jaeger et~al.(2004)Jaeger, Surkova, Blagov, Janssens, Kosman, Kozlov
  et~al.}]{jaeger2004dynamic}
Jaeger J, Surkova S, Blagov M, Janssens H, Kosman D, Kozlov KN, et~al. 2004.
\newblock Dynamic control of positional information in the early Drosophila
  embryo.
\newblock Nature 430:368--371.

\bibitem[{Kalinka et~al.(2010)Kalinka, Varga, Gerrard, Preibisch, Corcoran,
  Jarrells, Ohler, Bergman and Tomancak}]{Kalinka:2010jt}
Kalinka AT, Varga KM, Gerrard DT, Preibisch S, Corcoran DL, Jarrells J, Ohler
  U, Bergman CM, Tomancak P. 2010.
\newblock Gene expression divergence recapitulates the developmental hourglass
  model.
\newblock Nature 468:811--4.
\newblock \doi{10.1038/nature09634}.

\bibitem[{Kaneko(2006)}]{kaneko2006life}
Kaneko K. 2006.
\newblock Life: an introduction to complex systems biology, volume 171.
\newblock Berlin: Springer.

\bibitem[{Kaneko(2007)}]{kaneko2007evolution}
Kaneko K. 2007.
\newblock Evolution of robustness to noise and mutation in gene expression
  dynamics.
\newblock PLoS One 2:e434.

\bibitem[{Levin et~al.(2012)Levin, Hashimshony, Wagner and
  Yanai}]{levin2012developmental}
Levin M, Hashimshony T, Wagner F, Yanai I. 2012.
\newblock Developmental Milestones Punctuate Gene Expression in the
  \textit{Caenorhabditis} Embryo.
\newblock Developmental cell 22:1101--1108.

\bibitem[{Masamizu et~al.(2006)Masamizu, Ohtsuka, Takashima, Nagahara,
  Takenaka, Yoshikawa, Okamura and Kageyama}]{masamizu2006real}
Masamizu Y, Ohtsuka T, Takashima Y, Nagahara H, Takenaka Y, Yoshikawa K,
  Okamura H, Kageyama R. 2006.
\newblock Real-time imaging of the somite segmentation clock: revelation of
  unstable oscillators in the individual presomitic mesoderm cells.
\newblock Proceedings of the National Academy of Sciences of the United States
  of America 103:1313--1318.

\bibitem[{Mjolsness et~al.(1991)Mjolsness, Sharp and
  Reinitz}]{mjolsness1991connectionist}
Mjolsness E, Sharp DH, Reinitz J. 1991.
\newblock A connectionist model of development.
\newblock Journal of theoretical Biology 152:429--453.

\bibitem[{M{\"u}ller(1869)}]{muller1869facts}
M{\"u}ller F. 1869.
\newblock Facts and arguments for Darwin.
\newblock London: John Murray.

\bibitem[{Murray(2002)}]{murray2002mathematical}
Murray JD. 2002.
\newblock Mathematical biology, volume~2.
\newblock Berlin: Springer.

\bibitem[{Pourqui{\'e}(2003)}]{pourquie2003segmentation}
Pourqui{\'e} O. 2003.
\newblock The segmentation clock: converting embryonic time into spatial
  pattern.
\newblock Science 301:328--330.

\bibitem[{Quint et~al.(2012)Quint, Drost, Gabel, Ullrich, B{\"o}nn and
  Grosse}]{Quint:2012uk}
Quint M, Drost HG, Gabel A, Ullrich KK, B{\"o}nn M, Grosse I. 2012.
\newblock A transcriptomic hourglass in plant embryogenesis.
\newblock Nature 490:98--101.
\newblock \doi{10.1038/nature11394}.

\bibitem[{Richardson and Keuck(2002)}]{richardson2002haeckel}
Richardson MK, Keuck G. 2002.
\newblock Haeckel's ABC of evolution and development.
\newblock Biological Reviews 77:495--528.

\bibitem[{Salazar-Ciudad et~al.(2001{\natexlab{a}})Salazar-Ciudad, Newman and
  Sol{\'e}}]{Salazar-Ciudad:2001lq}
Salazar-Ciudad I, Newman SA, Sol{\'e} RV. 2001{\natexlab{a}}.
\newblock Phenotypic and dynamical transitions in model genetic networks. {I}.
  Emergence of patterns and genotype-phenotype relationships.
\newblock Evol Dev 3:84--94.

\bibitem[{Salazar-Ciudad et~al.(2001{\natexlab{b}})Salazar-Ciudad, Sol{\'e} and
  Newman}]{Salazar-Ciudad:2001db}
Salazar-Ciudad I, Sol{\'e} RV, Newman SA. 2001{\natexlab{b}}.
\newblock Phenotypic and dynamical transitions in model genetic networks. {II}.
  Application to the evolution of segmentation mechanisms.
\newblock Evol Dev 3:95--103.

\bibitem[{Sander and Schmidt-Ott(2004)}]{sander2004evo}
Sander K, Schmidt-Ott U. 2004.
\newblock Evo-Devo aspects of classical and molecular data in a historical
  perspective.
\newblock Journal of Experimental Zoology Part B: Molecular and Developmental
  Evolution 302:69--91.

\bibitem[{Soyer(2012)}]{soyer2012evolutionary}
Soyer OS. 2012.
\newblock Evolutionary systems biology, volume 751.
\newblock Berlin: Springer.

\bibitem[{ten Tusscher and Hogeweg(2011)}]{Ten-Tusscher:2011vn}
ten Tusscher KH, Hogeweg P. 2011.
\newblock Evolution of networks for body plan patterning; interplay of
  modularity, robustness and evolvability.
\newblock PLoS Comput Biol 7:e1002208.
\newblock \doi{10.1371/journal.pcbi.1002208}.

\bibitem[{Turing(1952)}]{am1952chemical}
Turing AM. 1952.
\newblock The chemical basis of morphogenesis.
\newblock Philosophical Transactions of the Royal Society of London Series B,
  Biological Sciences .

\bibitem[{von Baer(1828)}]{von1828ueber}
von Baer KE. 1828.
\newblock Ueber Entwicklungsgeschichte der Thiere.
\newblock Beobachtung und Reflexion 3.

\bibitem[{von Dassow et~al.(2000)von Dassow, Meir, Munro and
  Odell}]{von2000segment}
von Dassow G, Meir E, Munro EM, Odell GM. 2000.
\newblock The segment polarity network is a robust developmental module.
\newblock Nature 406:188--192.

\bibitem[{Waddington(1957)}]{waddington1957strategy}
Waddington C. 1957.
\newblock The strategy of the genes: a discussion of some aspecs of theoretical
  biology.
\newblock London: Allen and Unwin.

\bibitem[{Wagner(2005)}]{wagner2005robustness}
Wagner A. 2005.
\newblock Robustness and evolvability in living systems.
\newblock Princeton, NJ: Princeton University Press.

\bibitem[{Wang et~al.(2013)Wang, Pascual-Anaya, Zadissa, Li, Niimura, Huang,
  Li, White, Xiong, Fang, Wang, Ming, Chen, Zheng, Kuraku, Pignatelli, Herrero,
  Beal, Nozawa, Li, Wang, Zhang, Yu, Shigenobu, Wang, Liu, Flicek, Searle,
  Wang, Kuratani, Yin, Aken, Zhang and Irie}]{Wang:2013fq}
Wang Z, Pascual-Anaya J, Zadissa A, Li W, Niimura Y, Huang Z, Li C, White S,
  Xiong Z, Fang D, Wang B, Ming Y, Chen Y, Zheng Y, Kuraku S, Pignatelli M,
  Herrero J, Beal K, Nozawa M, Li Q, Wang J, Zhang H, Yu L, Shigenobu S, Wang
  J, Liu J, Flicek P, Searle S, Wang J, Kuratani S, Yin Y, Aken B, Zhang G,
  Irie N. 2013.
\newblock The draft genomes of soft-shell turtle and green sea turtle yield
  insights into the development and evolution of the turtle-specific body plan.
\newblock Nat Genet 45:701--6.
\newblock \doi{10.1038/ng.2615}.

\end{thebibliography}
\end{document}